\documentclass[twocolumn,trackchanges]{aastex701}

\usepackage{soul} 
\usepackage{amsmath}
\usepackage{xspace}
\usepackage{xifthen}
\usepackage{eso-pic}

\definecolor{forestgreen}{HTML}{228B22}
\definecolor{urlblue}{HTML}{000000}

\mathchardef\mhyphen="2D

\newlength{\dhatheight}

\newcommand{\unit}[1]{\ensuremath{\mathrm{\,#1}}\xspace}

\newcommand{\Msun}{\unit{M_\odot}}

\newcommand{\e}{\unit{e^{-}}}

\newcommand{\figref}[1]{Figure~\ref{fig:#1}}

\newcommand{\bandvar}[2][]{%
  \ifthenelse{\isempty{#1}}{\var{#2}}{\var{#2\_#1}}%
}

\newcommand{\LCDM}{\ensuremath{\rm \Lambda CDM}\xspace}

\newcommand{\var}[1]{\ensuremath{\texttt{\MakeUppercase{#1}}}\xspace}

\providecommand\physrep{\ref@jnl{Phys.~Rep.}}%
\providecommand\apjs{\ref@jnl{ApJS}}%
\providecommand{\jcap}{\ref@jnl{JCAP}}%

\usepackage{comment}
\usepackage{placeins}


\defcitealias{riley_auriga_2025}{AS1}
\defcitealias{shipp_auriga_2025}{AS2}

\begin{document}

\title{Hidden Disruption in Milky Way-like Satellite Galaxy Populations}
\shorttitle{Hidden Disruption}
\shortauthors{Thoron et al.}

\author[0009-0009-0252-5808]{Gray Thoron}
\affiliation{Department of Astronomy and DiRAC Institute, University of Washington, Seattle, WA 98195, USA}
\email{gthoron@uw.edu}

\author[0000-0003-2497-091X]{Nora Shipp}
\affiliation{Department of Astronomy and DiRAC Institute, University of Washington, Seattle, WA 98195, USA}
\email{nshipp@uw.edu}

\author[0000-0001-5805-5766]{Alexander~H.~Riley}
\affiliation{Lund Observatory, Division of Astrophysics, Department of Physics, Lund University, SE-221 00 Lund, Sweden}
\affiliation{Institute for Computational Cosmology, Department of Physics, Durham University, South Road, Durham DH1 3LE, UK}
\email{alexander.riley@fysik.lu.se}

\author[0000-0002-8354-7356]{Arpit Arora}
\affiliation{Department of Astronomy and DiRAC Institute, University of Washington, Seattle, WA 98195, USA}
\email{arora125@uw.edu}

\author[0000-0001-9985-1814]{Christine~M.~Simpson}
\affiliation{Argonne Leadership Computing Facility, Argonne National Laboratory, Lemont, IL 60439, USA}
\email{csimpson@anl.gov}

\author[0000-0002-4554-4488]{Rebekka Bieri}
\affiliation{University of Zurich, Department of Astrophysics,  Winterthurerstrasse 190, 8057 Zurich, Switzerland}
\email{rebekka.coles-bieri@uzh.ch}

\author[0000-0001-6957-1627]{Peter S. Ferguson}
\affiliation{Department of Astronomy and DiRAC Institute, University of Washington, Seattle, WA 98195, USA}
\email{pferguso@uw.edu}

\author[0000-0002-4350-7632]{Jaclyn Jensen}
\affiliation{Department of Physics \& Astronomy, Dartmouth College, 6127 Wilder Laboratory, Hanover, NH 03755, USA}
\email{jaclyn.R.Jensen@dartmouth.edu}

\author[0000-0003-4232-8584]{Facundo A. Gómez}
\affiliation{Departamento de Astronomía, Universidad de La Serena, Av. Juan Cisternas 1200 Norte, La Serena, Chile}
\email{fagomez@userena.cl}

\author[0000-0001-9667-1340]{Robert J. J. Grand}
\affiliation{Astrophysics Research Institute, Liverpool John Moores University, 146 Brownlow Hill, Liverpool L3 5RF, UK}
\email{r.j.grand@ljmu.ac.uk}

\author[0000-0003-3308-2420]{Rüdiger Pakmor}
\affiliation{Max-Planck-Institut für Astrophysik, Karl-Schwarzschild-Str. 1, D-85748 Garching, Germany}
\email{rpakmor@mpa-garching.mpg.de}

\author[0000-0002-6301-638X]{Freeke {van de Voort}}
\affiliation{Cardiff Hub for Astrophysics Research and Technology, Department of Physics and Astronomy, School of Physical, Chemical and Environmental Sciences, Cardiff University, Queen’s Buildings, Cardiff 14 CF24 3AA, UK}
\email{vandevoortf@cardiff.ac.uk}

\correspondingauthor{Gray Thoron}
\email{gthoron@uw.edu}

\begin{abstract}
Determining whether a satellite galaxy is tidally disrupting from photometric observations alone remains a fundamental challenge in near-field cosmology. Cosmological simulations of Milky Way-mass galaxies predict that satellite populations should be highly disrupted, yet the observed Milky Way satellite population appears largely intact. Understanding this tension requires either uncovering disruption that has gone undetected in Milky Way satellites, or identifying limitations in the simulations themselves. We use the Auriga suite of cosmological simulations to evaluate several commonly used indicators of disruption. We find that, in the simulated satellite population, the shape, size, and proper motion alignment of a satellite's bound component are unreliable indicators of its present-day disruption state despite their widespread use in the literature. In contrast, the extended stellar distribution, specifically the ratio of stellar mass between $3r_{1/2}$ and $10r_{1/2}$ to that within $r_{1/2}$, is strongly correlated with disruption. We introduce the extended mass ratio $\mu_{\rm EMR}$ as a metric for comparing extended stellar distributions across satellites of varying masses and exploring their connection to disruption in Milky Way observations. Applying this metric to a sample of Milky Way satellites with measured extended stellar distributions, we find that many are consistent with significant ongoing disruption, even among systems traditionally classified as intact. These results suggest that tidal disruption may be substantially more widespread in the observed Milky Way satellite population than currently recognized, and establish extended stellar distributions as a promising new tool for uncovering this hidden disruption.
\end{abstract}


\section{Introduction}
\label{sec:intro}

Cosmological baryonic simulations are among the most powerful tools for studying galaxy formation and evolution \citep{Feldmann_Bieri_2026}. Within a \LCDM framework, such simulations reproduce observations of large-scale structure, galaxy clusters, and individual galaxies \citep{sales_baryonic_2022, somerville_physical_2015, springel_simulations_2005}. In these simulations, structures assemble hierarchically -- dark matter halos merge over time, and stars and gas follow the underlying potential \citep{white_core_1978, white_galaxy_1991}. Galaxies form within the dark matter halos and subsequently merge. Smaller galaxies can orbit a more massive host galaxy as satellites, and are disrupted by the tidal field of the host galaxy \citep{bullock_tracing_2005, johnston_tracing_2008, cooper_galactic_2010}. This process of tidal disruption produces features such as streams and shells \citep{deason_eating_2016, sanderson_reconciling_2018, monachesi_auriga_2019, fattahi_tale_2020, vera-casanova_linking_2022} that may persist in the stellar halo of the host galaxy.

Satellite galaxies provide some of the most powerful tests of \LCDM on small scales. Early comparisons between observed and simulated populations of low-mass galaxies revealed a number of apparent discrepancies \citep[see][for a review]{bullock_small-scale_2017}, which have largely been resolved by simulations with higher resolution and more complete baryonic physics \citep{wetzel_reconciling_2016, garrison-kimmel_not_2017, fitts_fire_2017, simpson_quenching_2018, kim_missing_2018, buck_nihao_2019, sales_baryonic_2022} together with deeper and wider surveys of the Milky Way (MW) halo \citep{collaboration_eight_2015, drlica-wagner_milky_2020, tan_ultra-faint_2026}. Current simulations of MW-mass galaxies reproduce the abundance, radial distribution, and internal properties of the observed satellite population reasonably well. However, recent comparisons of disrupting satellites and stellar streams reveal a new discrepancy: these same simulations predict far more tidal disruption than is observed among the MW satellites \citep{shipp_streams_2023, riley_auriga_2025, shipp_auriga_2025}. This raises the question of whether simulated satellite populations match the real MW as well as their successes elsewhere suggest.

\citet{riley_auriga_2025}, hereafter Auriga Streams 1 (\citetalias{riley_auriga_2025}), found roughly twice as many massive streams as intact satellites of similar mass around simulated MW-mass hosts, considering systems with $M_* \geq 5\times 10^5$\Msun. In contrast, the MW has only a sixth as many known streams as intact satellites above this mass limit \citep{pace_local_2025, li_s_2022}. Similarly, the Feedback In Realistic Environments (FIRE-2) suite of zoomed cosmological-baryonic simulations of MW-mass galaxies \citep{wetzel_public_2023, wetzel2025second} shows roughly four times as many massive streams as are observed around the MW \citep{panithanpaisal_galaxy_2021, arora2022stability, horta2023observable}. In \citet{shipp_auriga_2025}, hereafter Auriga Streams 2 (\citetalias{shipp_auriga_2025}), the disruption of satellite galaxies is closely tied to orbital history, with smaller pericenters correlated with greater tidal disruption. While the overall distances and pericenters are consistent with the MW, the disruption in the Auriga simulations can occur at larger distances than it does within the MW: the simulated streams have pericenters up to roughly three times the largest pericenter of a MW stream and have much larger apocenters. Meanwhile, intact MW satellites exist at far smaller distances than their simulated counterparts. As with previous small-scale structure discrepancies, the apparent disagreement in disruption rates is likely shaped by both simulation and observational limitations.

Subsequent studies have attempted to resolve this discrepancy through more direct comparisons. \citet{shipp_streams_2023} created mock Dark Energy Survey (DES) observations of streams and massive satellites in the FIRE-2 simulations and found that, when accounting for observational completeness, the number of detected streams is consistent with simulations. The extended tidal features were often too faint to be observed with current telescopes, and many stream progenitors would be misclassified as intact satellite galaxies, suggesting that tidal disruption may be widespread among the apparently intact satellite population.

Because many stream features are predicted to be too faint for current telescopes, attention has turned to the stars surrounding the bright, detectable cores of satellites. In photometric observations without velocity information, a satellite lacking a visible tidal tail or other disruption feature is typically considered intact, and disruption must be inferred from structural parameters -- the 2D projected morphology (shape) and physical extent (size) -- of the apparently undisrupted component. In large-scale surveys with radial velocity or proper motion measurements -- for example from \textit{Gaia} -- individual member stars can be kinematically associated with their progenitor satellite even at large projected distances, revealing the extended stellar component. Tidal disruption analyses combine these approaches along with backward orbit integration to quantify the orbit and the tracks of possible tidal debris. 

Satellites that are known to be disrupting often have tidal tails: Crater II has tidal tails extending beyond 10 half-light radii $r_{1/2}$ \citep{atzberger_decam_2026, limberg_s5_2025, vivas_is_2026}, Tucana III has clear tidal tails \citep{collaboration_eight_2015, shipp_stellar_2018, li_first_2018}, and Sagittarius has a full stellar stream \citep{ibata_dwarf_1994, newberg_ghost_2002, majewski_two_2003, vasiliev_breath_2020}. Bo\"otes III, often associated with the Styx stream, is likely tidally disrupted, though its connection to the Styx stream remains uncertain \citep{carlin_bootes_2018,li_bootes_2026, jensen_bootes_2026}. \citet{chiti_extended_2021} identified members of Tucana II out to $\sim 9r_{1/2}$, and subsequent membership analyses have identified stars well beyond the central bound region of $\sim$12 MW satellites, in some cases reaching out to $12r_{1/2}$ \citep{jensen_small-scale_2024, tau_extended_2024}. Antlia II has a very large half-light radius of 3 kpc for its small stellar mass ($\sim10^6 \Msun$) \citep{torrealba_hidden_2019, ji_kinematics_2021}. Even satellite galaxies believed to be intact have spectroscopically confirmed members at large distances: Hercules has at least three member stars at $r>3r_{1/2}$ \citep{ou_signatures_2024, longeard_pristine_2022} and Ursa Minor has five at $r>5r_{1/2}$ \citep{sestito_extended_2023}. However, it remains uncertain whether such extended stellar distributions are the product of tidal effects or accreted stellar halos. With limited information available for most satellite systems, determining whether structural parameters and extended stellar distributions are reliable indicators of disruption is essential.

Ongoing and upcoming surveys -- including deeper photometry and improved proper motions from the Vera C. Rubin Observatory's Legacy Survey of Space and Time \citep[LSST;][]{ivezic_lsst_2019} 
and the Nancy Grace Roman Space Telescope \citep{spergel_wfirst-24_2013}, improved proper motions from future \textit{Gaia} data releases \citep{gaia_collaboration_gaia_2016}, additional spectroscopy from the Dark Energy Spectroscopic Instrument \citep[DESI;][]{cooper_overview_2023}, narrow-band photometry from MAGIC \citep{Chiti_MAGIC_2026}, as well as the \textit{Euclid} mission \citep{euclid_collaboration_euclid_2025}, 4MOST \citep{de_jong_4most_2019}, WEAVE \citep{dalton_weave_2012}, and the ARRAKIHS mission\footnote{\href{https://www.cosmos.esa.int/documents/7423467/7423486/ESA-F2-ARRAKIHS-Phase-2-PUBLIC-v0.9.2.pdf/61b363d7-2a06-1196-5c40-c85aa90c2113?t=1667557422996}{Arrakihs Phase 2 Proposal (public, pdf)}} -- will likely uncover new tidal disruption features around MW satellites. However, even these next-generation facilities will not detect all disruption signatures: tidal tails remain low surface brightness, and the faintest satellites will continue to be characterized primarily by their high-surface-brightness bound components. Interpreting both present and future observations therefore requires understanding what the structural properties of the bound component can -- and cannot -- tell us about disruption. The answer matters well beyond classifying individual systems: dwarf galaxy scaling relations, the satellite mass function, and the dynamical masses that underpin dark matter constraints all rest on the assumption that the observed satellites are intact, equilibrium systems. If tidal disruption is widespread among the apparently intact population, these inferences inherit a significant, unrecognized systematic.

In this paper, we use the Auriga simulations \citep{grand_auriga_2017, grand_overview_2024} to investigate whether observable properties -- projected sizes, shapes, and extended stellar distributions -- can reliably distinguish disrupting from intact systems. Because the simulations provide the true disruption state of every satellite, we can follow how these properties evolve as satellites are stripped, from infall to present day, and test each proposed indicator against the full population. Although we do not construct mock observations, we measure 2D projected quantities analogous to those available observationally, rather than the 3D quantities typically used in simulation studies. This paper is organized as follows. Section~\ref{sec:methods} describes the simulations and our methods. Section~\ref{sec:structure_sims} examines structural parameters and their evolution through disruption. Section~\ref{sec:emr} presents the extended stellar distribution analysis and introduces the extended mass ratio. Section~\ref{sec:mwcomp} compares our results to observed MW satellites. Section~\ref{sec:discussion} discusses our findings.

\section{Methodology}\label{sec:methods}

To evaluate observable indicators of disruption, we require a sample of satellites with known disruption states. We use the Auriga simulations for this purpose, as they allow us to directly calculate the gravitational bound fraction of each satellite while simultaneously measuring its observable structural properties. In this section we outline our methods. The Auriga simulations are described in Section~\ref{sec:sims}. The satellite and stream catalog is from \citetalias{riley_auriga_2025} and \citetalias{shipp_auriga_2025} and is described in Section~\ref{sec:sats}. We define the bound component of each satellite in Section~\ref{sec:bound}, and describe our structural parameter fitting in Section~\ref{sec:profile}.

\subsection{Auriga Simulations}\label{sec:sims}
The Auriga simulations are a suite of cosmological magnetohydrodynamic zoom-in simulations of 30 isolated MW-mass ($1$--$2 \times 10^{12}\,\Msun$) halos \citep{grand_auriga_2017, grand_overview_2024}. All 30 halos were simulated at the fiducial ``level 4'' resolution, six were resimulated at the higher ``level 3'' resolution, and one was resimulated at the highest ``level 2'' resolution. We use the six level 3 halos and their satellite populations, which have an initial baryon mass resolution of $6.7 \times 10^{3}\,\Msun$, a dark matter particle mass of $3.6 \times 10^{4}\,\Msun$, and a maximum softening length of 188\,pc \citep{grand_overview_2024}. The level 3 simulations have 63 snapshots, with a time cadence of $\sim 300$\,Myr near $z=0$.

The Auriga halos were selected from the EAGLE dark-matter-only simulations with a comoving side length of 100\,Mpc. The simulations used the Planck Collaboration cosmological parameters $\Omega_M = 0.307$, $\Omega_\Lambda= 0.693$, $h = 0.6777$, $\sigma_8 = 0.8288$, and $n_s = 0.9611$, and the initial conditions were created using PANPHASIA \citep{jenkins_new_2013}. The Auriga simulations were performed using the moving-mesh code \textsc{AREPO}, incorporating primordial and metal-line cooling \citep{vogelsberger_model_2013}, spatially uniform redshift-dependent UV background for reionization \citep{faucher-giguere_new_2009}, star formation and stellar feedback with an effective equation of state for the interstellar medium \citep{springel_cosmological_2003, vogelsberger_model_2013}, magnetic fields \citep{pakmor_magnetic_2017}, and seeding, accretion, and feedback from black holes \citep{springel_modelling_2005}.

The simulations produce realistic spiral disk galaxies with similar stellar mass and size to the MW \citep{grand_auriga_2017, marinacci_properties_2017}, and host realistic populations of satellite galaxies that in many ways resemble the MW's own satellite population \citep{simpson_quenching_2018, monachesi_auriga_2019}. These similarities have enabled robust comparisons, including determinations of the MW's assembly history, satellite orbits, and total mass \citep{deason_slight_2017, fattahi_origin_2019, riley_velocity_2019, callingham_mass_2018, deason_local_2019}, making Auriga well suited for studying the properties of intact and disrupting satellites around MW-like hosts.

We center and rotate the coordinate system of each simulation on the host halo at each snapshot. The rotation matrix is defined such that the $z$-axis is aligned with the principal angular momentum of the young stellar disk, defined as star particles $\le 1$ Gyr old and within $\le 8$ kpc of the galactic center at present day.

\subsection{Disrupting Satellites}\label{sec:sats}
We use the catalog of accreted satellites presented in \citetalias{riley_auriga_2025}. This sample is derived from the subhalo catalogs \citep{grand_auriga_2017}, which were constructed using the on-the-fly \texttt{SUBFIND} halo finder \citep{springel_populating_2001}. The host halo is identified using a Friends-of-Friends (FoF) algorithm applied to the dark matter particles \citep{davis_evolution_1985}. The \texttt{SUBFIND} algorithm then identifies subhalos within the host halo by determining whether particles are gravitationally bound to their subhalo. Each star particle is assigned to the subhalo of its nearest dark matter particle. The FoF and \texttt{SUBFIND} algorithms are applied across every snapshot and a merger tree constructed with the \texttt{LHaloTree} algorithm \citep{springel_simulations_2005} connects the individual \texttt{SUBFIND} halos and FoF groups across snapshots.

We associate all dark matter and star particles with their progenitor halo by tracing them back to the subhalo from which they were accreted onto the host galaxy, identified at its peak mass. All stars and dark matter particles belonging to each accretion event are then tracked to present day, regardless of whether they remain gravitationally bound to their progenitor halo or not. Due to resolution limits, we only track accretion events with at least 100 star particles at infall, corresponding to a lower mass limit of $M_* = 5 \times 10^5 \Msun$, leaving 404 accretion events (hereafter satellites) across the six host galaxies \citep{riley_auriga_2025}.

\begin{figure*}
    \centering
    \includegraphics[width=0.8\linewidth]{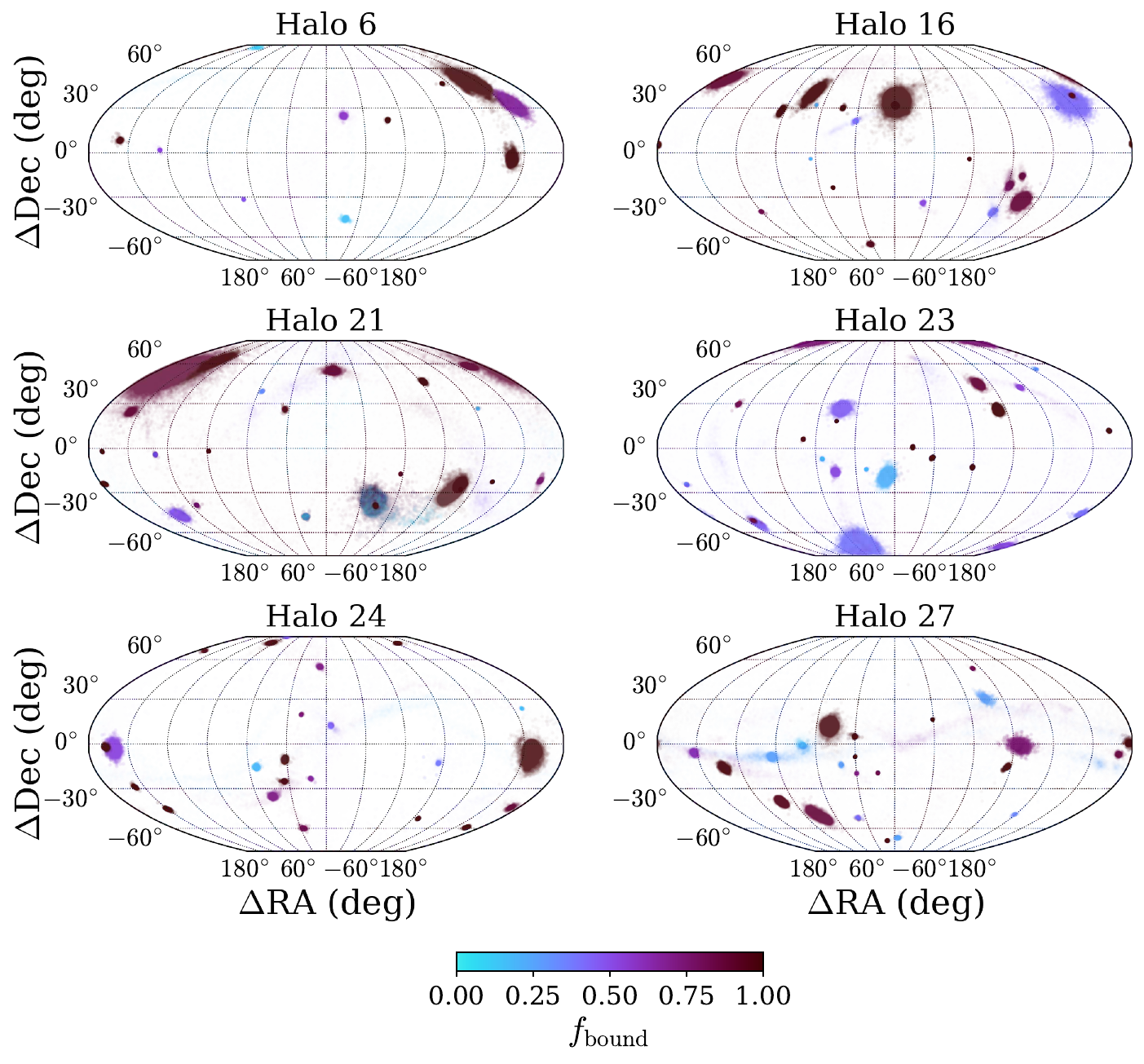}
    \caption{The on-sky distribution of satellites, as observed from the center of the host, and their tidal debris in each halo at present day, colored by the fraction of star particles that remain gravitationally bound to the satellite. Less disrupted satellites are shown at higher opacity to highlight their structure; mostly disrupted satellites are shown at lower opacity or omitted for clarity. Even largely intact satellites, at a range of distances, can have extended features stretching well across the sky.}
    \label{fig:fullsky}
\end{figure*}

\begin{figure}
    \centering
    \includegraphics[width=\columnwidth]{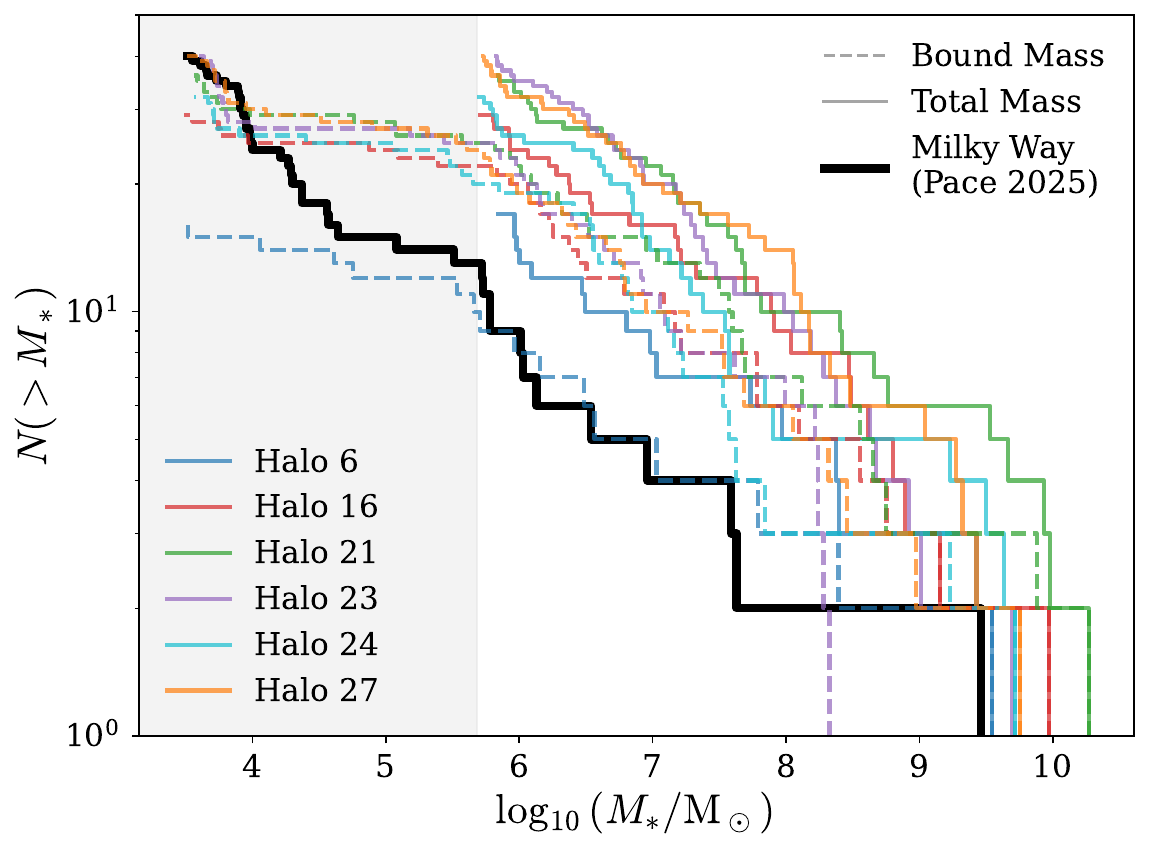}
    \caption{The satellite population in each halo as a function of stellar mass compared to the MW. Colored solid lines show the total stellar mass of each accreted satellite; dashed lines show the gravitationally bound mass at present day. The black solid line is the MW mass function from \citet{pace_local_2025}. The shaded region ($M_* \leq 10^{5.7}$ \Msun) indicates our resolution limit (100 star particles). Satellites in our sample may have a bound mass below this limit if a fraction of their star particles have been tidally stripped.}
    \label{fig:mass_pop_func}
\end{figure}

The population of accreted satellite galaxies of each host is shown in on-sky coordinates in Figure~\ref{fig:fullsky}. The color and opacity of each satellite reflect its bound fraction (as described in Section~\ref{sec:bound}). Mostly disrupted satellites are omitted for clarity. There is clear variation in satellite populations between hosts -- for example, halo 21 has a large number of accretion events including several massive mergers, while halo 6 has a relatively quiet merger history. Figure~\ref{fig:mass_pop_func} shows the cumulative mass distributions of the satellite populations. The total stellar mass function of the accreted systems is significantly different from the tidally evolved mass function of the same accreted systems. Colored solid lines show the total stellar mass of each accreted satellite; dashed lines show the gravitationally bound mass at present day. The solid black line is the MW population from \citet{pace_local_2025}. Only Auriga halo 6 has a comparable satellite population to the MW, while the other host halos feature similar slopes with larger overall populations. For masses lower than $5\times 10^5$ \Msun (the grey region), our sample consists only of satellites with a peak mass above this limit that have since disrupted.

\subsection{The Bound Component}\label{sec:bound}

The goal of this work is to study how the properties of the bound components of accreted satellites evolve as systems undergo tidal disruption. We determine which stars remain bound to their progenitor and which have been tidally stripped. We define the \textit{bound component} to be the particles associated with the satellite that are gravitationally self-bound. The rest of the satellite -- all stars and dark matter no longer gravitationally bound, including tidal tails and phase-mixed debris -- makes up the \textit{disrupted component}. 

We identify the bound and disrupted components of each accreted satellite across all host galaxies and all simulation snapshots from accretion to present day. Previous work has used the bound fractions reported in \texttt{SUBFIND}. We instead use an iterative unbinding method, as \texttt{SUBFIND} has been shown to under-identify small-radius subhalos at moderate resolutions \citep{mansfield_symfind_2024}, and because an iterative unbinding method can be applied consistently across simulation suites \citep[e.g.][]{MandacaruGuerra:2026}. Bound fractions calculated with our iterative unbinding method are comparable to those from \texttt{SUBFIND} as presented in \citetalias{riley_auriga_2025, shipp_auriga_2025} at high and low bound fractions. The iterative unbinding method recovers higher bound fractions than \texttt{SUBFIND} at intermediate bound fractions, indicating that our method more reliably traces the mass evolution of satellites experiencing high levels of disruption. A detailed comparison is presented in Appendix~\ref{app:fbound}.

Our iterative unbinding method considers particles to be bound to the progenitor if their total energy satisfies $E = v^2/2 + V(r) \leq 0$, where $v$ is the velocity and $V(r)$ is the potential energy due to the self-gravity of the satellite. The self-gravity of each satellite is calculated using only star and dark matter particles. We exclude gas and black holes from this calculation; dark matter is the dominant mass component of these systems and stars are the primary observational tracer. At the mass scales considered here, gas fractions in dwarf satellites are negligible, so this exclusion does not meaningfully affect our results.
The potential energy of the dark matter and star particles in each satellite is calculated using a multipole Poisson solver from \texttt{Agama} \citep[Action-based Galaxy Modeling Architecture;][]{vasiliev_agama_2019}\footnote{The potential energy of the dark matter and star particles is robust with respect to the choice of using a tree-based or multipole Poisson potential solver.}. To determine the center of the gravitational potential, we fit a mass-weighted kernel density estimate (KDE) to the positions and velocities of the associated star particles and take the center to be the mean position of particles within the highest 1\% of the KDE. We then iteratively recompute the potential and update the set of bound particles until convergence, defined as a change of less than $0.1\%$ in the number of bound particles. If fewer than 5 bound particles remain, the satellite is considered fully disrupted.

\begin{figure}
    \centering
    \includegraphics[width=\linewidth]{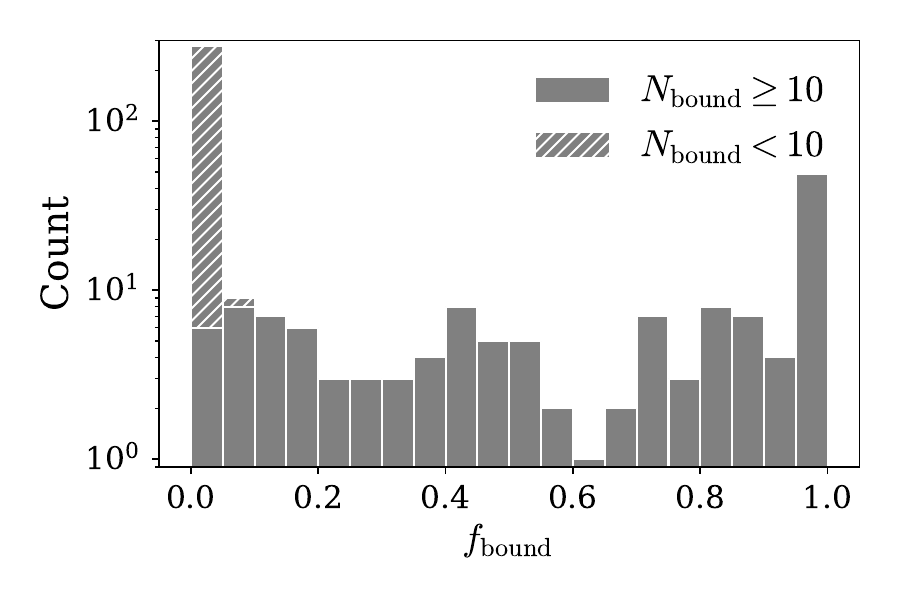}
    \caption{The present-day distribution of bound fractions. Satellites with 10 or more bound star particles (solid histogram) constitute the sample analyzed throughout this work; those with fewer than 10 bound particles are shown as a hatched histogram for reference. Our sample spans the full range of disruption levels.}
    \label{fig:fbound_hist}
\end{figure}

Once we determine the bound and disrupted components of each satellite, we calculate the bound fraction at a given snapshot $f_{\rm bound}$, defined as the total mass of gravitationally bound star particles ($M_{*,\text{bound}}$) divided by the total stellar mass associated with the satellite ($M_{*,\text{total}}$):
\begin{equation}
    f_{\rm bound}(t) = \frac{M_{*,\text{bound}}(t)}{M_{*,\text{total}}(t)}.
\end{equation}
$f_{\rm bound} = 1$ indicates a completely intact satellite, while $f_{\rm bound} = 0$ corresponds to a completely disrupted satellite. We show the present-day distribution of stellar bound fractions in Figure~\ref{fig:fbound_hist}. We refer to the stellar bound fraction as ``the bound fraction'' throughout the rest of this work. We omit satellites with fewer than ten bound star particles, resulting in a present-day sample of 141 satellites. The lowest bound fraction we find is $f_{\rm bound} \sim 0$, with only 18 out of 31440 particles bound. For $f_{\rm bound} \leq 0.1$ our sample is incomplete; however, we are primarily interested in systems that retain a significant bound component, so this does not affect our conclusions.

\subsection{Fitting Structural Parameters}\label{sec:profile}
MW satellite galaxies are primarily discovered and characterized via deep photometric imaging, which is most sensitive to the bright, centrally concentrated stellar distribution. The structural parameters most easily measurable from these data are the half-light radius $r_{1/2}$ (the satellite's ``size'') and ellipticity $\varepsilon$ (the satellite's ``shape''), which are typically reported in catalogs such as the Local Volume Database \citep{pace_local_2025}. These are measured by fitting member stars to an elliptical density profile, often a 2D Plummer profile \citep{plummer_problem_1911}, assuming a constant mass-to-light ratio. We adopt this method to calculate the structural parameters of the simulated satellite galaxies.

Unlike in observations, we can determine the entire gravitationally bound component of each satellite and restrict our sample to stars that remain bound to the satellite. We fit a Plummer profile to this bound component, without weighting by luminosity or mass.

\begin{figure}
    \centering
    \includegraphics[width=\linewidth]{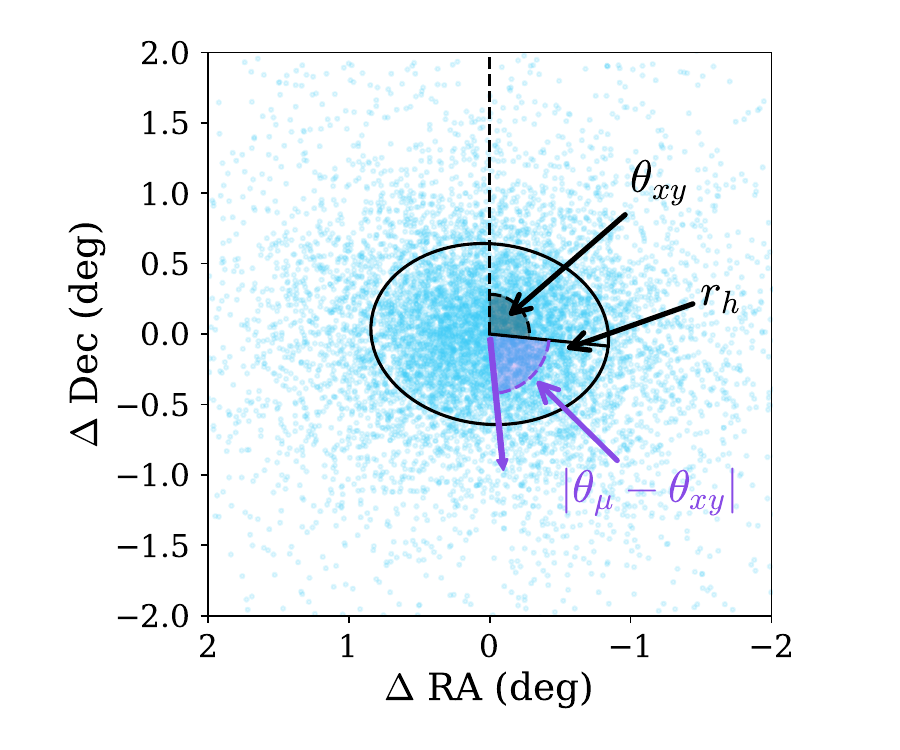}
    \caption{An example elliptical Plummer profile fit (black ellipse) to the on-sky stellar distribution (blue points) of a simulated satellite. The position angle $\theta_{xy}$ defines the rotation of the ellipse. $r_{h}$ (black line) is the semi-major axis, related to $r_{1/2}$ by $r_{1/2} = r_h \sqrt{1-\varepsilon}$. The angle between the proper motion (purple arrow) and the major axis, $|\theta_{\mu} - \theta_{xy}|$, defines the alignment between the proper motion and the direction of elongation.}
    \label{fig:plummer_model}
\end{figure}

Prior to fitting, the bound component of each satellite is projected onto a tangent plane centered on the dwarf, giving projected coordinates $(X, Y)$ (these projected coordinates are akin to right ascension and declination). The surface density of a 2D elliptical Plummer profile is
\begin{equation}
    \Sigma(R) = \frac{M}{\pi}\frac{r_h^2}{\left(r_h^2+R^2\right)^2}
    \label{eq:plummerprofile}
\end{equation}
where $M$ is the total projected mass, $r_{h}$ is the Plummer scale radius, and $R$ is the elliptical radius. To compute $R$, we define offsets from the fitted center $(\bar{X}, \bar{Y})$,
\begin{align}
    \Delta X &= X - \bar{X}, \qquad \Delta Y = Y - \bar{Y},
\end{align}
rotate into the major- and minor-axis frame by the position angle $\theta_{xy}$,
\begin{align}
    X' &= \Delta X \sin\theta_{xy} + \Delta Y \cos\theta_{xy}, \\
    Y' &= -\Delta X \cos\theta_{xy} + \Delta Y \sin\theta_{xy},
\end{align}
and form
\begin{equation}
    R = \left[ X'^2 + \frac{Y'^2}{(1-\varepsilon)^2} \right]^{1/2},
    \label{eq:elliptical_radius}
\end{equation}
where $\theta_{xy}$ is the position angle of the major axis measured clockwise from the $Y$-axis and $\varepsilon \in [0, 1)$ is the ellipticity, defined as $\varepsilon = 1 - b/a$ for major and minor axes $a$ and $b$. The azimuthally averaged projected half-light radius is related to the Plummer scale radius by
\begin{equation}
    r_{1/2} = r_h \sqrt{1-\varepsilon}.
    \label{eq:half_light_radius}
\end{equation}
An example of the ellipse defining this profile is shown in \figref{plummer_model} in black. The background blue points are on-sky positions of star particles and the purple arrow points in the direction of the proper motion of the satellite. The proper motion used throughout this paper is equivalent to the proper motion of a star corrected for solar reflex motion. 

We use a maximum likelihood estimator to determine the best-fit parameters $\{\bar{X},\ \bar{Y},\ r_h,\ \varepsilon,\ \theta_{xy}\}$ as implemented in the ultra-faint galaxy likelihood \verb|ugali| toolkit package \citep{bechtol_ugali_2015, collaboration_eight_2015}, which was developed for and has been widely applied to MW satellite galaxies. For each parameter, we sample the posterior with MCMC and report the median of a one-dimensional kernel density estimate, with uncertainties given by the 68\% highest density interval. 

Observations cannot determine which stars are gravitationally bound, so the structural parameters of observed satellites are measured from member samples that may include stripped stars. To test whether our energy-based selection biases the comparison, we refit all satellites using radial membership cuts (Appendix~\ref{app:members}). The structural parameters from the energy-based selection agree with those from a cut at $R \leq 3r_{1/2}$ for $f_{\rm bound} \gtrsim 0.3$, and diverge only for the most disrupted systems, where larger cuts include unbound stars. Differences in membership selection therefore cannot account for the structural differences between the Auriga and MW satellites discussed in Section~\ref{sec:mwcomp}. 

\subsection{Numerical Resolution}\label{sec:resolution}

Following the tidal evolution of low-mass satellites is a demanding numerical problem, and cosmological hydrodynamical simulations at the resolution of Auriga are sensitive to several numerical effects: finite force resolution and discreteness noise can enhance tidal mass loss \citep{van_den_bosch_dark_2018, van_den_bosch_disruption_2018, errani_can_2020, chiang_universal_2025}, and the sizes and shapes of simulated galaxies depend on the gravitational softening and can be inflated by spurious collisional heating between particles of unequal mass \citep{ludlow_energy_2019, ludlow_numerical_2020}. Comparisons of the same Auriga halos simulated at different resolutions find that bound fractions and stellar morphologies are well converged, with a modest tendency for satellites to retain more of their stars at higher resolution \citepalias{riley_auriga_2025}, and an Auriga simulation with dark matter particles comparable in mass to the star particles yields satellite disruption times consistent with the reference simulation \citep{pakmor_superstars_2025}. 
We use the level 3 satellite population, the highest-resolution sample of Auriga satellites available across multiple hosts, consistent with previous studies of the Auriga satellite population, including \citetalias{riley_auriga_2025} and \citetalias{shipp_auriga_2025}. As in any cosmological simulation, the level 3 Auriga runs are not free of numerical effects, particularly for the lowest-mass satellites, and we assess which satellites in our sample may be affected using commonly used criteria based on force resolution and particle discreteness in Appendix~\ref{app:resolution}. The convergence of bound fractions across resolution levels supports the disruption states used throughout this work, while the detailed structural evolution of disrupting satellites has not yet been tested across resolution; we discuss the implications in Section~\ref{sec:disc_resolution} and leave a detailed characterization of these effects to future work.

\section{Structure and Evolution of Disrupting Satellites}
\label{sec:structure_sims}
We now examine whether the structural parameters of the bound component -- half-light radius $r_{1/2}$, ellipticity $\varepsilon$, and position angle $\theta_{xy}$ -- can reliably distinguish disrupting from intact satellites. We begin with three example satellites spanning a range of present-day bound fractions to build intuition for how these parameters evolve, then turn to the full population to assess whether systematic trends hold across the diverse satellite histories in Auriga.

\subsection{Example Satellites}
\label{sec:examples}
\begin{figure*}
    \centering
    \includegraphics[width=0.8\linewidth]{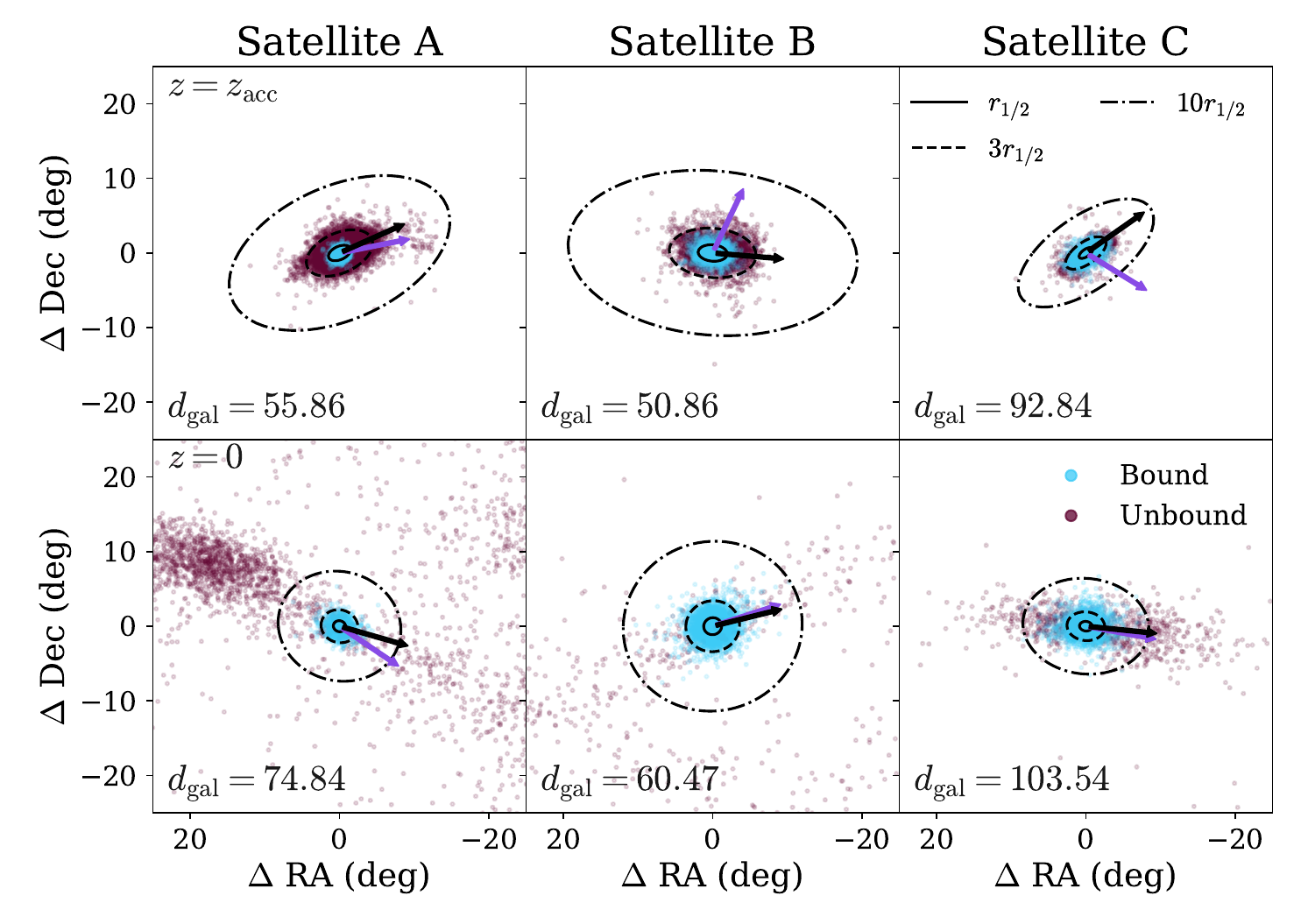}
    \caption{The fits of Satellite A (left), Satellite B (center), and Satellite C (right). The top row shows the fit at the time of accretion; the bottom row shows the fit at present day. The solid ellipse is $r_{1/2}$; dashed is $3 r_{1/2}$; and dot-dash is $10 r_{1/2}$. Stars in the bound component at present day are in cyan, while stars in the unbound component are in dark red. The purple arrow denotes the direction of proper motion, and the black arrow shows the direction of elongation for the fit. While the bound components remain compact and well described by the elliptical fits, the unbound debris extends well beyond $10r_{1/2}$ at present day.}    \label{fig:acc_present_examples}
\end{figure*}

\begin{figure}
    \centering
    \includegraphics[width=\linewidth]{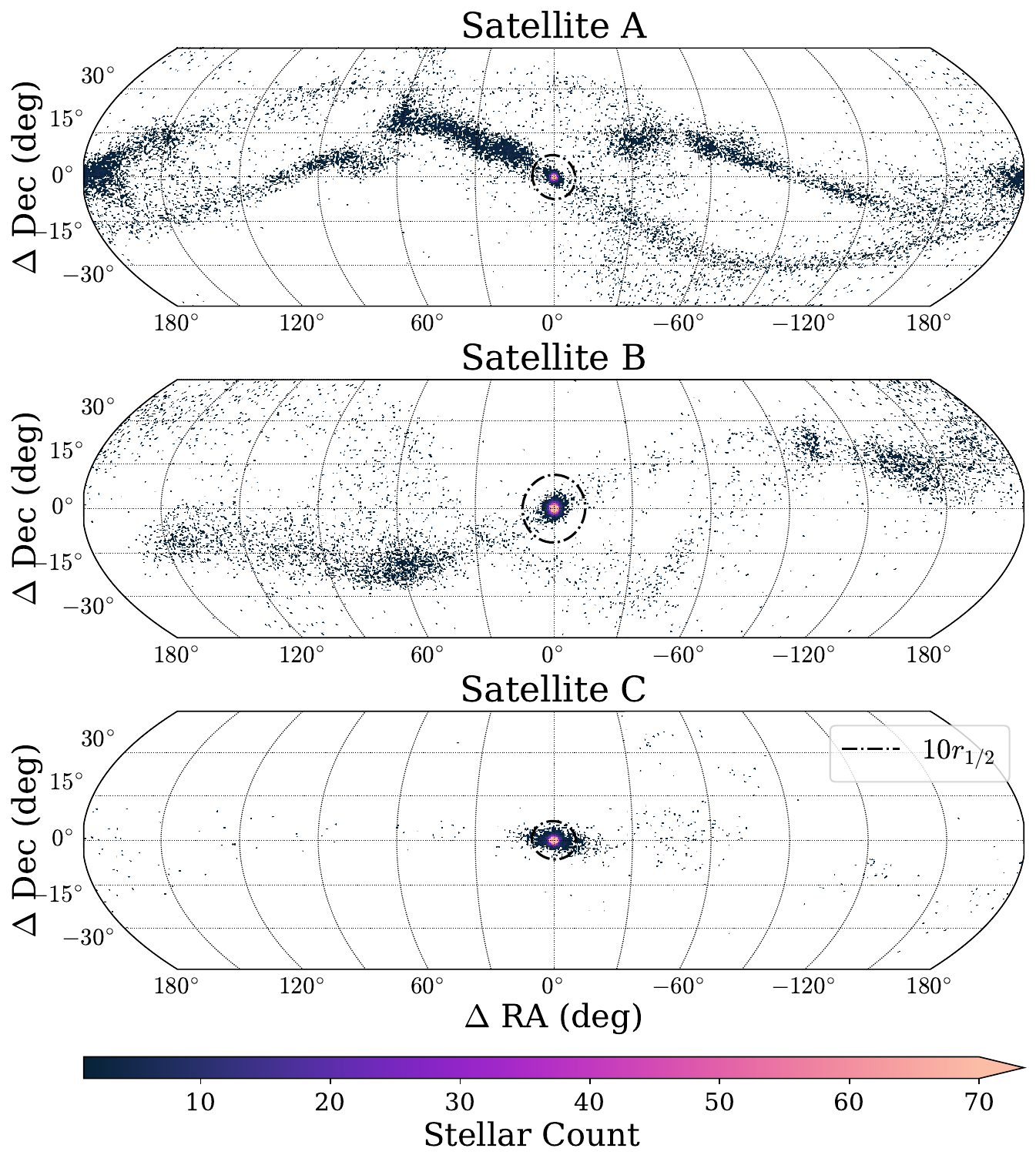}
    \caption{The on-sky projections across the entire sky for Satellites A (top), B (middle), and C (bottom). The stellar count is plotted as a hex-bin. The black dot-dashed ellipse is the fit with a radius of $10 r_{1/2}$. Even the most bound satellite (Satellite C; $f_{\rm bound} = 0.877$) has stars beyond $10 r_{1/2}$.}
    \label{fig:onsky_projs}
\end{figure}

To show the evolution of individual satellites, we present three sample satellites through time, which we refer to as Satellites A, B, and C. Satellite A is heavily disrupted at present day ($f_{\rm bound} = 0.135$), Satellite C is largely intact ($f_{\rm bound} = 0.877$), and Satellite B is intermediate ($f_{\rm bound} = 0.484$). Relevant properties of the satellites are presented in Table~\ref{tab:examples_properties}. Satellite A was chosen at random from the population of satellites with a low bound fraction at present day. Satellites B and C were chosen to have similar accretion times ($\sim 7.5$ Gyr ago) and masses to Satellite A ($\sim 10^8$ \Msun), but different present-day bound fractions. In practice, the available sample is limited: Satellites B and C are somewhat less massive than Satellite A, and Satellite B was accreted $\sim 9.5$ Gyr ago. We show the shape of the three sample satellites, zoomed in on their bound components, at present day and at the time of their individual accretions in Figure~\ref{fig:acc_present_examples}. Figure~\ref{fig:onsky_projs} shows the satellites' present-day morphologies across the whole sky. All satellites, including Satellite C, which has a bound fraction of $f_{\rm bound} = 0.877$, have bound stars out to $\sim 9r_{1/2}$. Throughout this work, we use Satellites A, B, and C as reference points for comparison to the full population.

\begin{deluxetable}{lccc}
\tablecaption{Properties of Example Satellites\label{tab:examples_properties}}
\tablewidth{0pt}
\tablehead{
\colhead{Property} & \colhead{Sat A} & \colhead{Sat B} & \colhead{Sat C}
}
\startdata
Host Halo & 16 & 21 & 21 \\
Satellite Peak Mass ID & 1148 & 4631 & 3746 \\
$M_{*,\mathrm{total}}$ ($10^7$\,$M_{\odot}$) & 10.9 & 6.5 & 4.2 \\
$M_{*,\mathrm{bound}}$ ($10^7$\,$M_{\odot}$) & 1.5 & 3.1 & 3.7 \\
$f_{\mathrm{bound}}$ & 0.135 & 0.484 & 0.877 \\
$t_{\mathrm{accretion}}$ (Gyr) & 7.65 & 9.51 & 7.99 \\
\enddata
\tablenotetext{}{Satellite A was chosen for its low present-day bound fraction; its bound component contains over 100 star particles, sufficient for reliable profile fitting. Satellites B and C were chosen to have accretion times and stellar masses close to those of Satellite A, but different present-day bound fractions; the available sample limits how closely the masses and accretion times were matched. Satellite B was accreted $\sim 9.5$ Gyr ago, $\sim 2$ Gyr earlier than Satellite A.}
\end{deluxetable}

\subsection{Structural Parameters at Present Day}
\label{sec:present_day}
A satellite's size at a given mass is one of the most commonly used photometric indicators of tidal disruption. on the expectation that tides strip mass while inflating the stellar distribution, lowering the density. However, as we show below, a larger size for a given mass is not a reliable indicator of disruption.

\begin{figure*}
    \centering
    \includegraphics[width=0.8\linewidth]{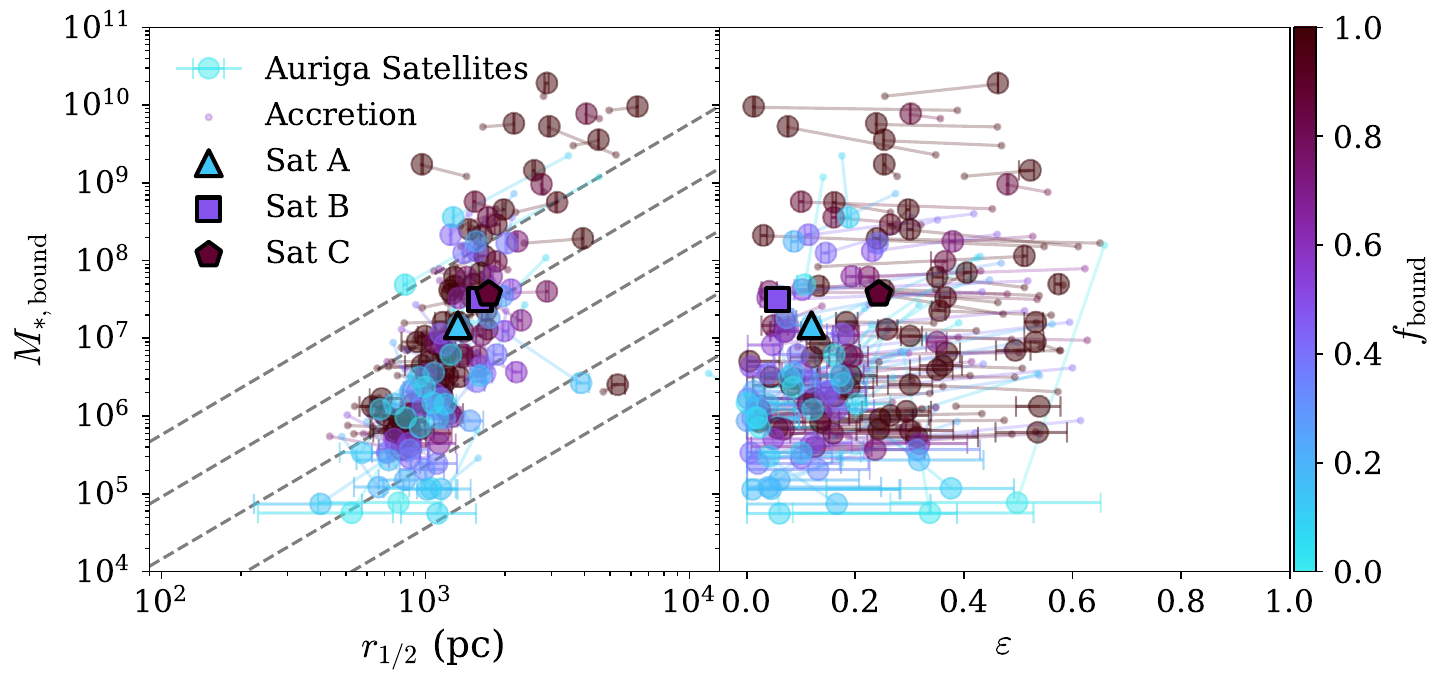}
    \caption{Left: Half-light radius $r_{1/2}$ versus present-day gravitationally bound mass, colored by bound fraction. The dashed grey lines are contours of constant surface brightness of ($24, 26, 28, 30, 32$) mag/arcsec$^2$ assuming a mass-to-light ratio of 2. Small points connected to larger points by a thin line show the location of each satellite at accretion. The mass-size relation shifts along a consistent track between accretion and present day. A mostly disrupted satellite is indistinguishable from a mostly intact one based on mass and size alone. Right: Ellipticity $\varepsilon$ versus present-day bound mass. Ellipticities are generally low across all bound fractions, with no correlation with disruption.}
    \label{fig:present_day_m_size_relation}
\end{figure*}

We show the mass-size relation of satellites at present day in Figure~\ref{fig:present_day_m_size_relation}. All satellites follow a single mass-size relation regardless of their bound fraction, with no clear separation between disrupted and intact systems. Similarly, there is no clear separation between intact and disrupted satellites in ellipticity. The more elliptical portion of our sample is overwhelmingly intact: only 8 out of 51 satellites with $\varepsilon > 0.2$ have $f_{\rm bound} \leq 0.5$. This does not imply that there is correlation between low ellipticity and disruption, as 45 of the 90 satellites with $\varepsilon \leq 0.2$ remain intact with $f_{\rm bound} > 0.5$, and no satellites have $\varepsilon > 0.6$ at present day. Both intact and disrupted satellites span a wide range of ellipticities, and the more elongated satellites are predominantly intact, so a satellite's ellipticity cannot be used to infer its disruption state.
A satellite that has lost most of its stellar mass is indistinguishable, in the mass-size plane, from an intact satellite that formed with the mass the stripped system has today. A single snapshot, however, cannot reveal whether these parameters respond to disruption as it proceeds; we therefore turn to their evolution from accretion to present day.

\subsection{Evolution of Structural Parameters}
\label{sec:evo_params}
\begin{figure*}
    \centering
    \includegraphics[width=0.7\linewidth]{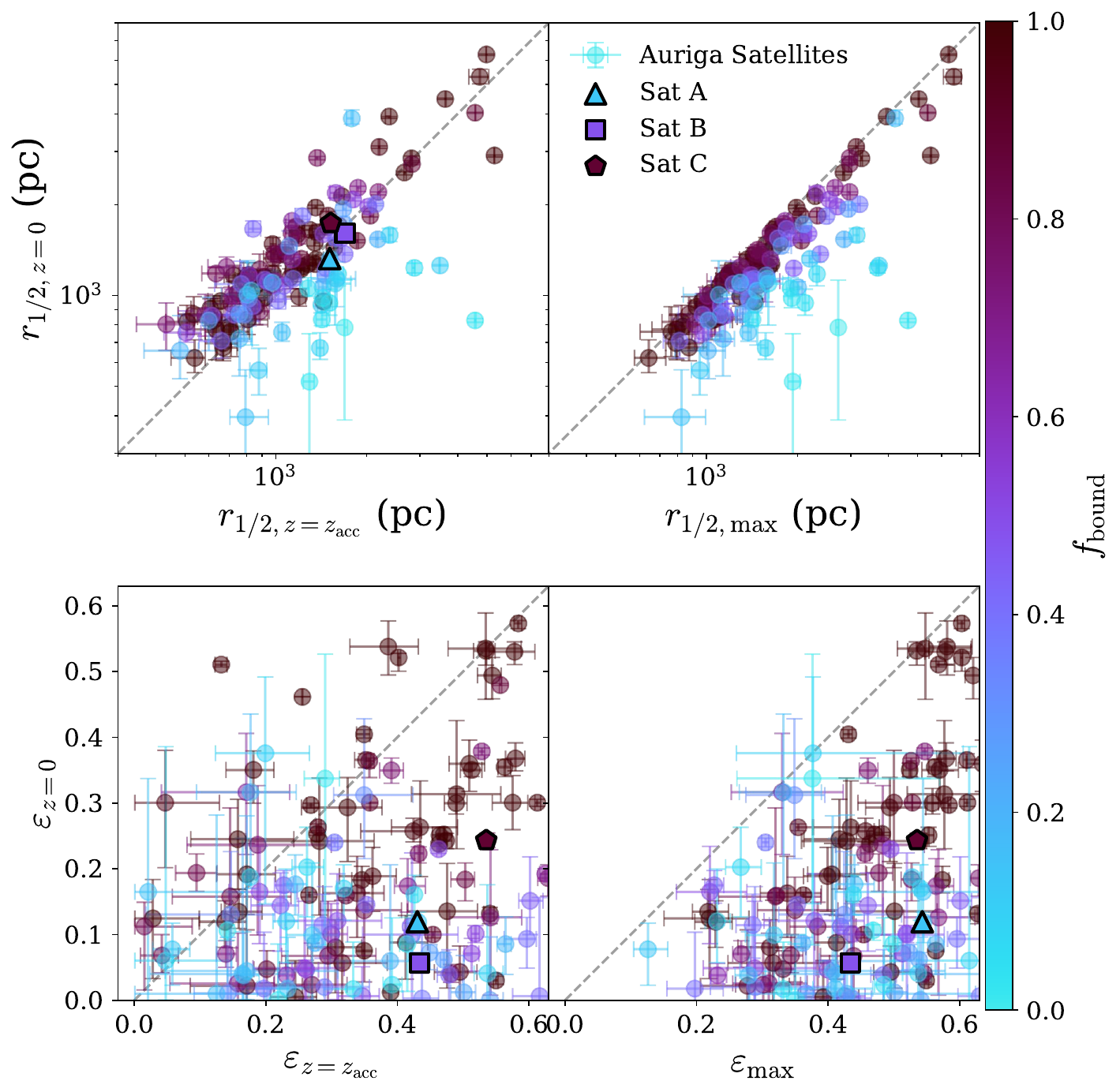}
    \caption{Left: the accretion-time versus present-day structural parameters, colored by stellar bound fraction at present day. Right: the maximum value of each structural parameter versus the present-day structural parameters. Top Row: half-light radius $r_{1/2}$. Bottom Row: ellipticity $\varepsilon$. The gray dashed line is the 1-to-1 line; satellites below the line have seen the parameter decrease since accretion, and those above have seen it increase. Satellites become rounder after accretion across the full range of bound fractions, and most become larger; only the most heavily disrupted systems fall substantially below the one-to-one line in size. Intact satellites are often still increasing in size, but all the disrupted satellites have reached their maximum size and are now decreasing in size. Proper motion alignment showed no trend with accretion-time parameters and is omitted.}
    \label{fig:evolvedparams}
\end{figure*}

At present day, sizes follow a clear mass-size relation while ellipticities, though scattered, remain small. Neither parameter shows a strong correlation with bound fraction. To understand how satellites of similar bound fractions have evolved over time, we compare present-day structural parameters to those at accretion. The structural properties of satellites at accretion closely resemble their present-day values: accreting satellites span the full range from compact to extended and from round to elliptical, and none of these parameters predict the present-day bound fraction, as shown in the left panels of Figure~\ref{fig:evolvedparams}. Both parameters do evolve systematically; however, neither evolves monotonically with mass loss. 
Satellites become rounder from accretion to present day across the full range of bound fractions; no satellite ever has an ellipticity above $0.6$. Satellites that retain nearly all of their stellar mass can be as circular as those that have lost most of it, with 45 of the 90 satellites with $\varepsilon \leq 0.2$ retaining $f_{\rm bound} > 0.5$, and Satellite C, with $f_{\rm bound} = 0.877$, drops from $\varepsilon = 0.54$ at accretion to $\varepsilon = 0.24$ today. Preferential stripping of stars in the outskirts leaves a rounder bound remnant and contributes to this trend, but cannot by itself explain circularization in satellites that have lost little mass. Tidal heating and the relaxation of systems that are dynamically disturbed at infall can contribute, as can spurious collisional heating, which can isotropize the stellar orbits of simulated galaxies independently of their tidal history \citep{ludlow_spurious_2021, wilkinson_impact_2023}; we discuss the possible numerical contribution in Section~\ref{sec:disc_resolution}.

Sizes evolve non-monotonically. Most satellites are larger at present day than at accretion, whether or not they have been substantially stripped. Comparing these two epochs alone obscures the full evolution: satellites first expand after infall, reach a maximum size, and then contract as tidal truncation of the outer profile takes over. A satellite's position along this track, measured as its present-day size relative to its maximum, correlates with bound fraction far more closely than its net change since accretion does, as shown in the top panels of Figure~\ref{fig:evolvedparams}.    Contraction of the stellar component under tidal stripping is the expected response seen in high-resolution idealized simulations, in which deeply embedded stellar components can also expand modestly before contracting \citep{errani_structure_2022}, and cosmological simulations of the Local Group find that stripping changes stellar sizes only modestly \citep{fattahi_tidal_2018}. In Auriga, however, we cannot determine how much of the initial expansion is tidal in origin, since spurious collisional heating by the more massive dark matter particles also inflates stellar systems over time, independently of their tidal history (Section~\ref{sec:disc_resolution}). The contraction phase, in contrast, is driven by stripping and is correlated with bound fraction. Satellites at their maximum size at present day are almost exclusively intact systems still on the rising branch, while those well below their peak are the most heavily stripped; many remain larger than at accretion even after contracting substantially from their peak, which is why comparison between present day and infall shows growth across nearly the whole population. Because both trends act on the population as a whole rather than in proportion to mass loss, they shift satellites along the mass-size relation rather than off it, and the present-day distributions of size and ellipticity look much the same for stripped and intact systems. We now examine the full evolutionary histories of our example satellites.

\begin{figure*}[t]
    \centering
    \includegraphics[width=\linewidth]{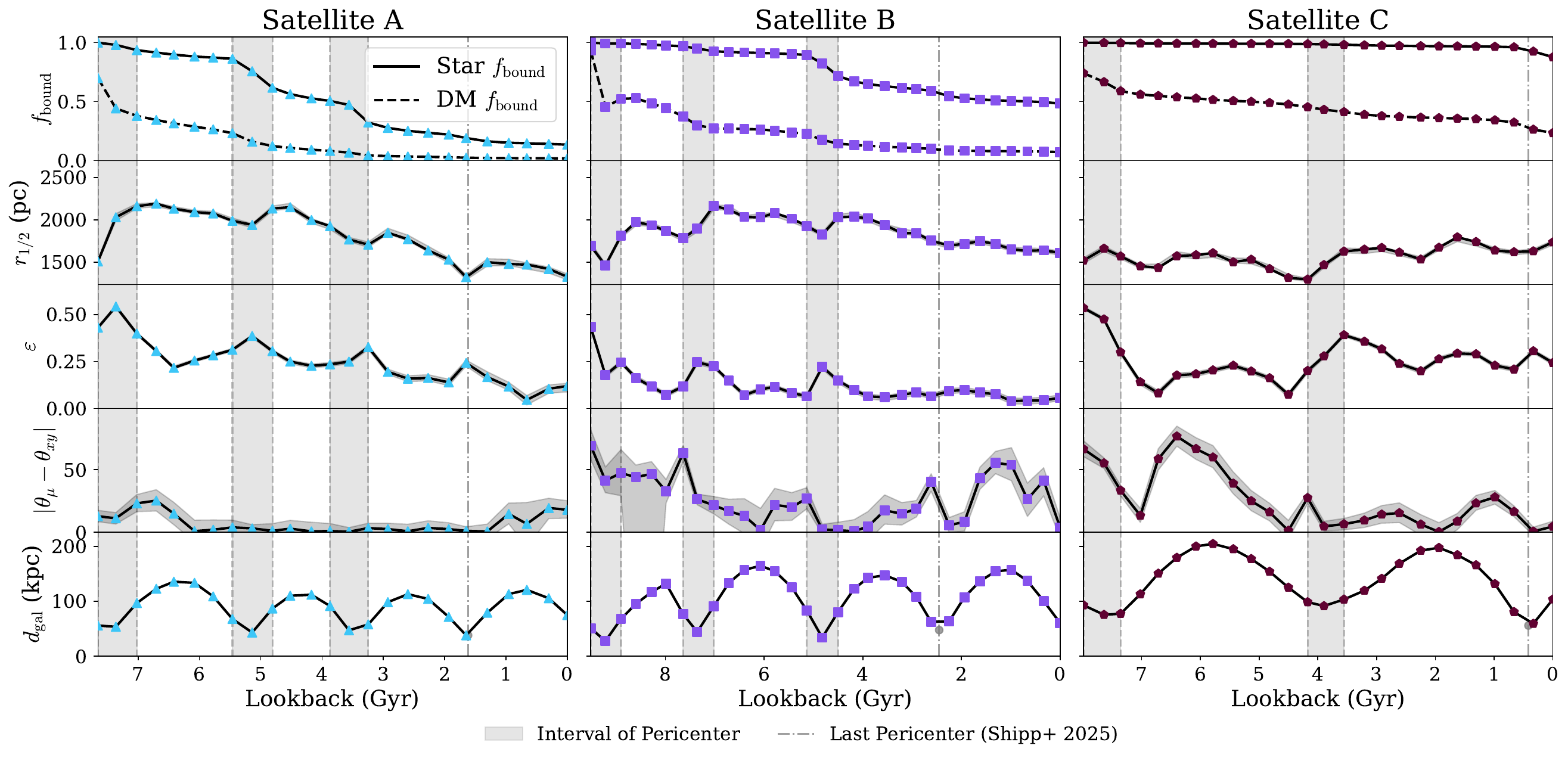}
    \caption{Structural evolution of Satellites A (Panel 1), B (Panel 2), and C (Panel 3) over lookback time. From top to bottom: bound fraction (solid: stellar, dashed: dark matter), half-light radius $r_{1/2}$, ellipticity $\varepsilon$, alignment between proper motion and direction of elongation, and distance from the host center. Markers show simulation output timesteps. Grey shaded vertical regions indicate pericentric passages; the grey dashed line marks the most recent pericenter from \citetalias{shipp_auriga_2025}, with a marker at the pericenter distance. Disruption occurs primarily at pericentric passages, but the structural parameters respond inconsistently: the pericentric peaks in ellipticity seen in Satellite A do not appear in Satellites B or C.}
    \label{fig:stream_properties}
\end{figure*}

Satellites show structural diversity at both accretion and present day. Figure~\ref{fig:stream_properties} shows the full evolutionary histories of Satellites A, B, and C with pericenters from \citetalias{shipp_auriga_2025}. For Satellite A (leftmost panel), the most disruption occurs at pericenter, where the ellipticity peaks and the half-light radius increases slightly before steadily decreasing; Satellite A is among the minority of systems that end up smaller than they began. This is consistent with the expectation that tidal forces are strongest at pericenter and with the transition from expansion to stripping-driven truncation described above. The proper motion and direction of elongation remain primarily aligned throughout, with small deviations uncorrelated with orbital phase. This behavior does not hold for Satellites B and C. The ellipticity and the half-light radius of both Satellites B and C peak at some pericenters, but this trend does not persist to the present day. At the last pericenter of Satellite B (center panel), the ellipticity and the half-light radius barely peak, and in Satellite C (rightmost panel) neither parameter shows a clear response to the most recent pericentric passage. Satellite C is the farthest from its host and experiences only two pericenters since accretion, while Satellites A and B experience three (see Appendix~\ref{app:gallery} for further examples).

These examples demonstrate that while orbital history strongly informs the rate of disruption, the structural parameters of the bound component do not reliably reflect it. Population-level trends exist, with satellites becoming rounder and, in most cases, larger after accretion. However, a satellite's present-day size and ellipticity depend far more on its initial structure and orbital history than on how much mass it has lost. At fixed bound fraction, the satellite-to-satellite scatter in these parameters swamps the systematic trends. The information that tracks disruption lives in a satellite's structural history, not in its present-day structure. Therefore, a present-day measurement of size or shape cannot be inverted into a disruption state. Having established that conventional structural parameters fail as disruption indicators, we turn to a different observable: the distribution of stars at large projected radii.

\section{Extended Stellar Distributions}
\label{sec:emr}

\begin{figure*}[t]
    \centering
    \includegraphics[width=\linewidth]{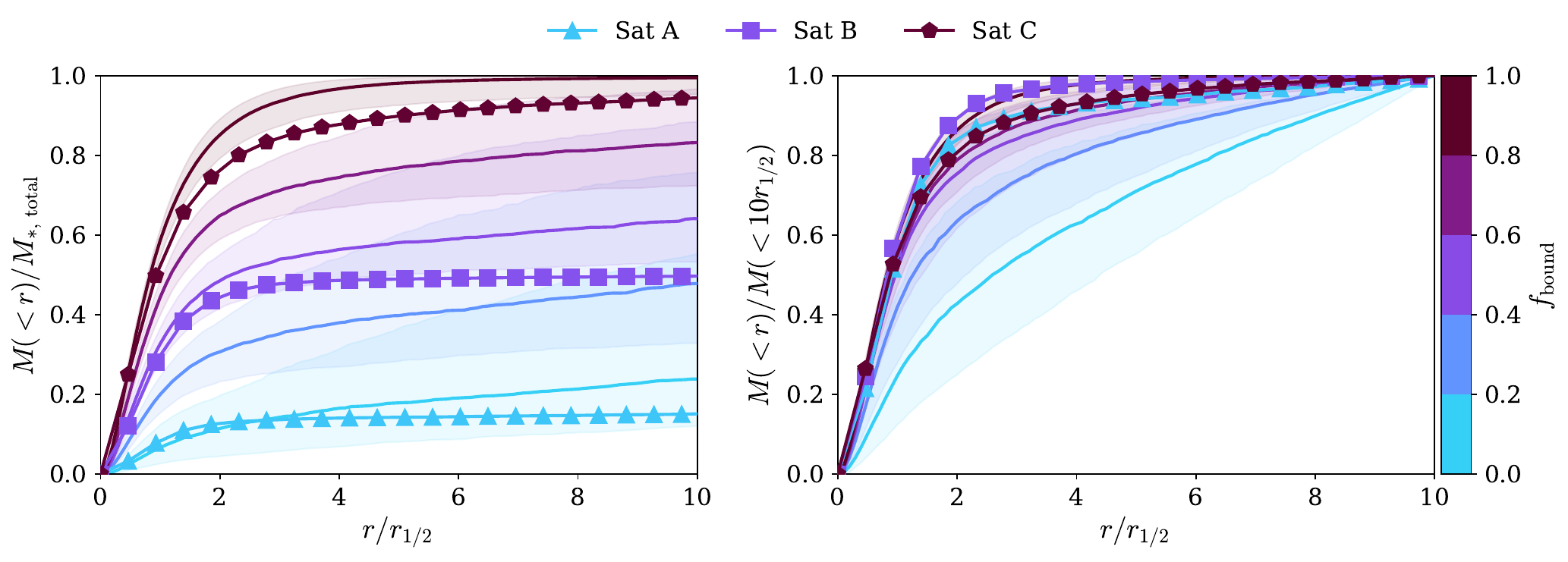}
    \caption{Cumulative radial stellar mass distributions, binned by bound fraction and normalized by total stellar mass (left) and by stellar mass within $10r_{1/2}$ (right). Five equally sized bound fraction bins are colored by bound fraction. Shaded regions show the 16th to 84th percentile interval; the solid lines show the mean within each bin. The solid lines with individual points correspond to the example satellites. The cumulative distributions show that the distribution of mass near the satellite galaxy varies drastically depending on its disruption state.}
    \label{fig:mass_profiles}
\end{figure*}

\begin{figure*}[t]
    \centering
    \includegraphics[width=0.9\linewidth]{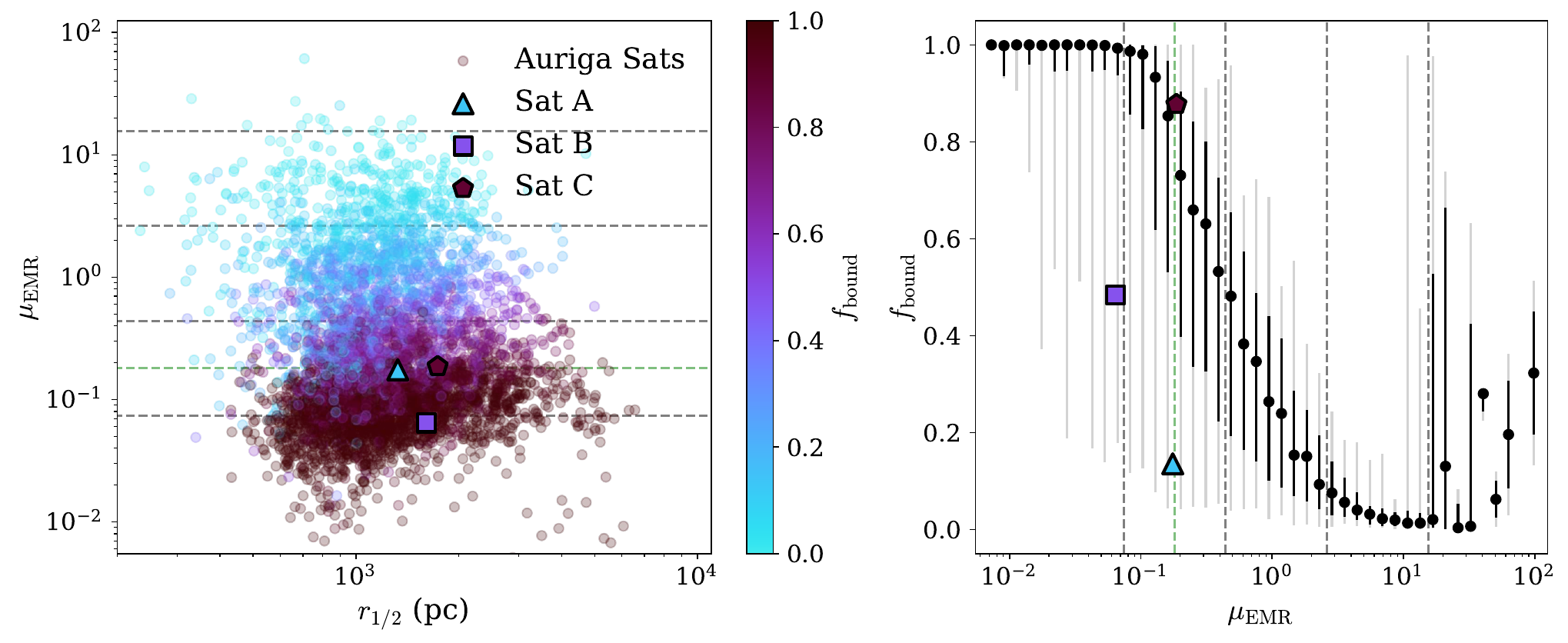}
    \caption{Left: $\mu_{\rm EMR}$ versus $r_{1/2}$ for all satellites at all snapshots since accretion, colored by bound fraction. Right: Median bound fraction as a function of $\mu_{\rm EMR}$. Black lines show the 16th and 84th percentile range; the solid grey line shows the full range within each bin. Dashed grey lines mark the bin edges from Table~\ref{tab:EMR_fbounds}. The dashed green line shows the expected $\mu_{\rm EMR}$ for an undisrupted Plummer profile ($\mu_{\rm EMR} \approx 0.182$). In both panels, a clear trend between $\mu_{\rm EMR}$ and the bound fraction of stars is visible.}
    \label{fig:M_ratio_halfrad}
\end{figure*}

While the structural parameters of the bound component are poor indicators of disruption, the distribution of stars at large projected radii offers a more direct window into a satellite's tidal history. Extended stellar distributions have been proposed as indicators of tidal disruption, and while they are low-surface-brightness features, their presence is increasingly detectable with recent advances in deep imaging, narrow-band photometry, precision astrometry, and wide-area spectroscopy \citep{jensen_small-scale_2024, tau_extended_2024, sestito_extended_2023, vera-casanova_linking_2022, longeard_pristine_2022, waller_cosmic_2022}. 

To quantify this signal, we define the extended mass ratio $\mu_{\rm EMR}$ as the ratio of stellar mass in the extended region ($3r_{1/2} < R < 10r_{1/2}$) to stellar mass in the inner region ($R < r_{1/2}$):

\begin{equation}
    \mu_{\rm EMR} = \frac{M(3r_{1/2} < R < 10r_{1/2})}{M(R < r_{1/2})}.
    \label{eq:emr}
\end{equation}

The inner boundary of $3r_{1/2}$ was chosen because the bound component is largely contained within this radius (as seen in Figure~\ref{fig:acc_present_examples}), while $10r_{1/2}$ captures the extended stellar distribution. We tested a variety of radial boundaries and found that this combination most clearly separates mostly bound from mostly disrupted satellites. A value of $\mu_{\rm EMR} = 0$ indicates no mass in the extended region. $\mu_{\rm EMR} = 1$ indicates that the extended and inner regions have the same mass. The inner region is defined by the half-light radius; it does not necessarily contain half the mass of the satellite galaxy.

Figure~\ref{fig:mass_profiles} shows the cumulative radial distribution of stellar mass in all satellites at all snapshots within our sample, including both bound and unbound star particles. Each shaded band shows the 16th and 84th percentiles across all satellites in each bound fraction bin, and the lines with points show the distributions of our three example satellites. There is a clear correlation with bound fraction: the most disrupted satellites have only $\sim 20\%$ of their total mass within $10r_{1/2}$, while the most intact satellites have nearly all their mass within this radius (left panel). Normalizing by mass within $10r_{1/2}$ (right panel) provides a more observationally motivated comparison. The profiles are more similar after normalization, yet differences between bound fraction bins persist at high levels of disruption ($f_{\rm bound} \leq 0.5$). The most disrupted satellites ($f_{\rm bound} < 0.2$) typically have the majority of the mass within $10r_{1/2}$ beyond $\sim 2r_{1/2}$. However, there is considerable variation within bins as demonstrated by Satellites A, B, and C, which have similar normalized radial profiles despite having different disruption levels.

Figure~\ref{fig:M_ratio_halfrad} (left panel) shows $\mu_{\rm EMR}$ versus half-light radius for all satellites at all snapshots since accretion. The most disrupted satellites tend to have $\mu_{\rm EMR} > 1$ while the most intact satellites have $\mu_{\rm EMR} < 0.1$, though there is considerable scatter at all values. As a reference, we plot a projected undisrupted satellite following a Plummer profile with $\varepsilon=0$. Such a satellite has $\mu_{\rm EMR} \approx 0.182$, meaning roughly one fifth as much mass in the extended region as in the inner region. Satellites with $\mu_{\rm EMR}$ near this value are consistent with $f_{\rm bound} \sim 0.80$, while satellites with significantly higher $\mu_{\rm EMR}$ have more mass in their extended envelopes than expected from their profile alone. Because an undisrupted Plummer profile can produce a similar $\mu_{\rm EMR}$ to a mildly disrupted satellite, a moderate $\mu_{\rm EMR}$ ($0.1\lesssim \mu_{\rm EMR} \lesssim 3$) alone cannot confirm significant disruption.

Figure~\ref{fig:M_ratio_halfrad} (right panel) shows the median bound fraction as a function of $\mu_{\rm EMR}$. The median bound fraction decreases with increasing $\mu_{\rm EMR}$; however, there is significant scatter at all values -- most bins contain satellites spanning a wide range of bound fractions, as illustrated by Satellites A and B. Table~\ref{tab:EMR_fbounds} quantifies this scatter, showing the bound fractions below which 10\%, 50\%, and 90\% of satellites in each bin are found. We discuss the implications of these findings for MW satellite observations in Section~\ref{sec:mwcomp}.

\begin{deluxetable}{lcccc}
\tablecaption{Bound Fraction Distribution as a Function of Extended Mass Ratio}
\tablewidth{0pt}
\tablehead{
\colhead{$\mu_\mathrm{EMR}$} &
\colhead{$N_{\rm sats}$} & 
\colhead{$f_{\mathrm{bound},10\%}$} & 
\colhead{$f_{\mathrm{bound},50\%}$} & 
\colhead{$f_{\mathrm{bound},90\%}$} 
}
\startdata
$[0, 0.074)$ & 1083 & 0.943 & 0.998 & 1.0 \\
$[0.074, 0.44)$ & 2631 & 0.499 & 0.857 & 0.994 \\
$[0.44, 2.62)$ & 1053 & 0.084 & 0.233 & 0.465 \\
$[2.62, 15.56)$ & 339 & 0.013 & 0.037 & 0.094 \\
$[15.56, \infty)$ & 36 & 0.002 & 0.036 & 0.593 \\
\enddata
\tablenotetext{}{Bound fraction percentiles for satellites in each $\mu_{\rm EMR}$ bin. The columns $f_{\mathrm{bound},10\%}$, $f_{\mathrm{bound},50\%}$, and $f_{\mathrm{bound},90\%}$ indicate the bound fraction values below which 10\%, 50\%, and 90\% of satellites are found, respectively. $N_{\rm sats}$ counts the total number of satellites in each bin across all hosts and timesteps.}
\label{tab:EMR_fbounds}
\end{deluxetable}

\section{Comparisons to Observed Satellites}
\label{sec:mwcomp}

We have established that, within our simulated sample of satellite galaxies, the extended stellar distribution is strongly correlated with disruption, while conventional structural parameters are not. We now compare our findings to observations of the MW satellite population.

\subsection{Extended Stellar Distributions of Satellites}
\label{sec:mwcomp_emr}

The ratio of stellar mass in the extended region ($3r_{1/2}$ to $10r_{1/2}$) to that within $r_{1/2}$ is strongly correlated with bound fraction, as shown in Figure~\ref{fig:M_ratio_halfrad}. Recent surveys have measured extended stellar distributions in MW satellites by searching for member stars beyond $r_{1/2}$. At least 12 MW satellites have member stars out to $\sim 12r_{1/2}$ \citep{jensen_small-scale_2024, tau_extended_2024, sestito_extended_2023}, and \citet{jensen_small-scale_2024} found extended distributions out to $\gtrsim 6.5r_{1/2}$ in nine out of 66 satellites.

\begin{figure*}[t]
    \centering
    \includegraphics[width=\linewidth]{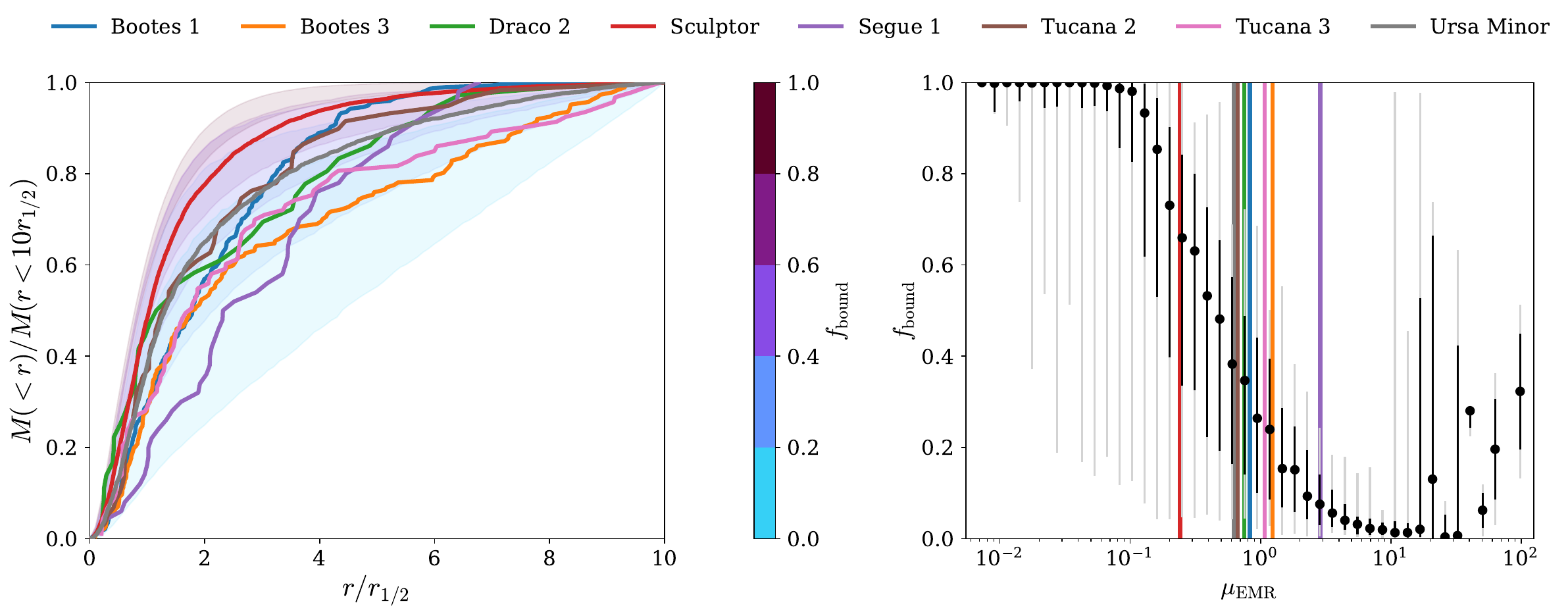}
    \caption{Comparison of extended stellar distributions between MW satellites and Auriga simulations. Left: cumulative stellar count profiles normalized to the total within $10r_{1/2}$, for eight MW satellites from \citet{jensen_small-scale_2024} (colored lines) alongside the Auriga bound fraction bins from Figure~\ref{fig:mass_profiles} (shaded bands). Right: $\mu_{\rm EMR}$ distribution from Figure~\ref{fig:M_ratio_halfrad} with the MW satellites shown as narrow vertical bands. All eight MW satellites have extended count ratios within the range populated by substantially disrupted Auriga satellites.}
    \label{fig:emr_MW_comparison}
\end{figure*}

We present a preliminary comparison between the Auriga satellites and eight\footnote{We exclude Grus II as it overlaps with the Orphan stream, which may contaminate the selection.} MW satellites from \citet{jensen_small-scale_2024} in Figure~\ref{fig:emr_MW_comparison}. The eight MW satellites are Bo\"otes I, Bo\"otes III, Draco II, Sculptor, Segue I, Tucana II, Tucana III, and Ursa Minor. The MW comparisons presented here are limited by observational completeness, and our values from simulations have no observational limits. This is an exercise to demonstrate the use of this metric. The left panel shows the cumulative stellar count normalized to the total within $10r_{1/2}$. The right panel shows the $\mu_{\rm EMR}$ distribution from Figure~\ref{fig:M_ratio_halfrad} alongside the stellar count ratios for the MW satellites, represented as vertical lines. All eight MW satellites have extended count ratios that fall within the range populated by significantly disrupted Auriga satellites. This is consistent with substantial tidal disruption, but does not establish the disruption state of any individual MW satellite. Satellites like Bo\"otes III and Tucana III have known tidal tails or streams associated with them \citep{grillmair_four_2009, carlin_bootes_2018, collaboration_eight_2015, shipp_stellar_2018}. The disruption of other satellites remains uncertain: the extended features of Draco II do not resemble tidal tails \citep{jensen_small-scale_2024}, and Bo\"otes I and Tucana II may have accreted a sub-satellite \citep{jensen_small-scale_2024, koposov_accurate_2011, chiti_detailed_2023}. Segue I has the highest extended count ratio, but lacks a complete stellar census out to $10r_{1/2}$, with its most distant member star at $6.76r_{1/2}$; if all stars within $r_{1/2}$ have been detected, this measurement acts as a lower bound on the extended count ratio.

Although $\mu_{\rm EMR}$ correlates strongly with bound fraction in Auriga, the scatter in this relation and differences between the simulated and observed populations prevent us from using it as a calibrated quantitative predictor of disruption for individual MW satellites. We highlight six caveats. First, a wide range of bound fractions can correspond to a similar $\mu_{\rm EMR}$, unless $\mu_{\rm EMR}\lesssim0.1$ or $\mu_{\rm EMR}\gtrsim 3$, making it difficult to predict the disruption level of any individual satellite. However, we find many measured ratios are consistent with disruption.

Second, we do not include a satellite galaxy's accreted stellar halo in the calculation of $\mu_{\rm EMR}$. Sub-satellites are treated as independent accretion events rather than components of their host satellite. In observations, the satellite's accreted stellar halo is often captured as part of the satellite, which would cause the satellite's disruption to be overestimated by $\mu_{\rm EMR}$.

Third, $\mu_{\rm EMR}$ is sensitive to the membership selection, which depends on the assumed spatial profile. \citet{jensen_small-scale_2024} modeled the spatial distribution with a two-component exponential profile; adjusting the profile shape or membership probability cutoff would affect the calculated ratio. The membership probability decreases with distance, and $\mu_{\rm EMR}$ is reliant on distant stars. Increasing the minimum membership probability decreases the maximum distance, which in turn decreases $\mu_{\rm EMR}$. A more conservative membership probability would result in a $\mu_{\rm EMR}$ more consistent with a higher $f_{\rm bound}$. 

\begin{figure*}[t]
    \centering
    \includegraphics[width=0.9\linewidth]{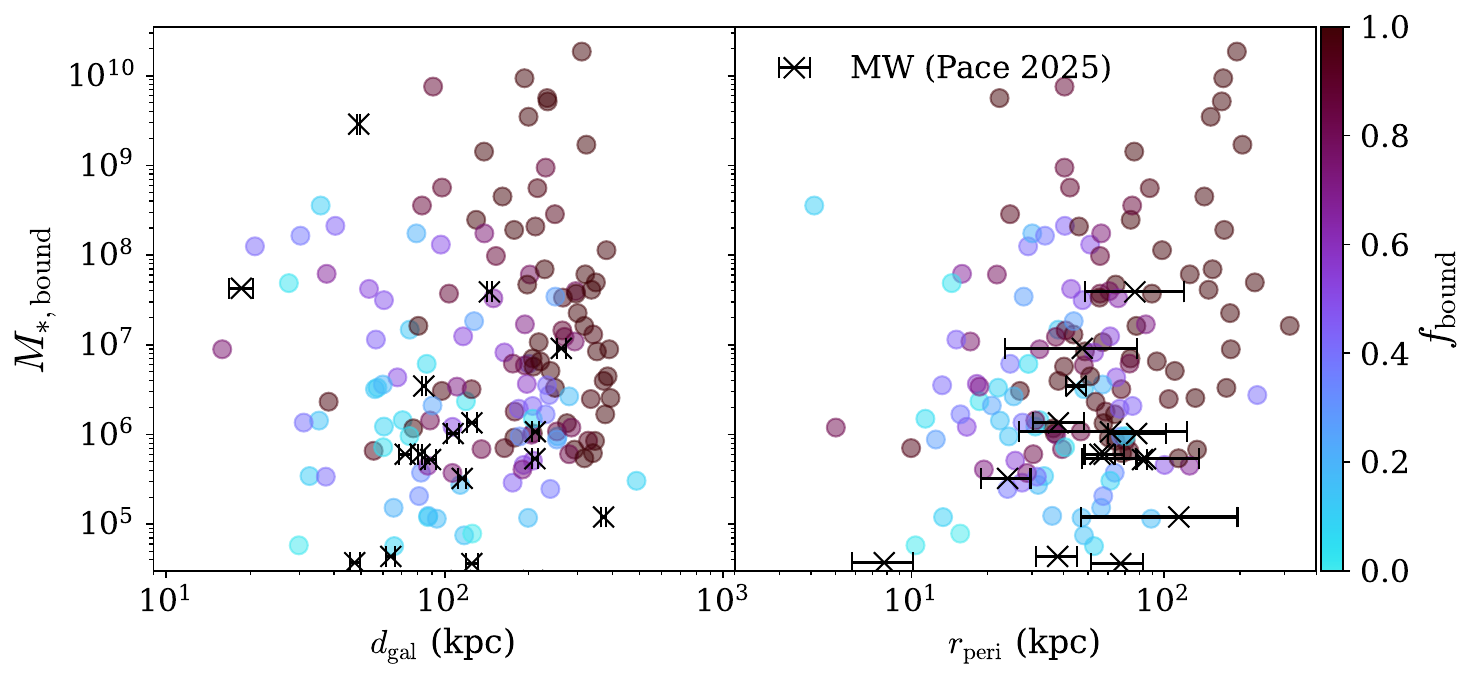}
    \caption{Left: Present-day Galactocentric distance $d_{\text{gal}}$ versus present-day gravitationally bound mass, colored by bound fraction. Black crosses show MW satellite galaxies. Right: Most recent pericentric distance $r_{\text{peri}}$ versus present-day bound mass. Auriga satellite pericenters are from \citetalias{shipp_auriga_2025}; MW satellite pericenters are from \citet{pace_proper_2022}. More disrupted satellites tend to have smaller present-day distances and pericenters.}
    \label{fig:orbit_MW_comp}
\end{figure*}

Fourth, the comparison spans a wider mass range than our simulated sample. Many of the observed satellites fall below $5 \times 10^5$ \Msun, the resolution limit of the Auriga catalog, and at these masses satellites are more strongly dark matter dominated than any system in our sample, which may alter their response to tidal stripping. We find that the correlation between $\mu_{\rm EMR}$ and bound fraction holds across the full mass range that we can resolve, but whether it extends to lower masses will need to be tested with higher-resolution simulations.

Fifth, $\mu_{\rm EMR}$ is defined as a stellar mass ratio in the simulations but is measured as a ratio of member star counts in the observations. These are equivalent only if the stellar mass function and the observational completeness do not vary with radius. Radial variation in either, for example from crowding in the satellite center or from a magnitude limit that samples different populations at different radii, would bias the observed ratio relative to the simulated one.

Sixth, the observed satellites in Figure~\ref{fig:emr_MW_comparison} are not a representative sample. They were selected from \citet{jensen_small-scale_2024} because they have detected member stars at large radii, so the fact that all of them have high extended count ratios does not tell us about the disruption state of the many MW satellites in which no extended stellar component has yet been detected. A population-level comparison will require a sample with uniform depth and membership criteria, or a forward model of the selection function.

Given these caveats, we do not infer bound fractions for individual MW satellites. Instead, we find that their extended stellar distributions are consistent with those of substantially disrupted satellites in Auriga, motivating further investigation of whether tidal disruption is more widespread in the MW than currently recognized. A quantitative comparison will require simulations resolving the full mass range of the observed population, together with forward modeling of the observational selection; we leave this to future work. Upcoming surveys will improve our ability to measure extended stellar distributions across a larger sample of MW satellites, extending to larger radii with cleaner membership samples.

\subsection{Satellite Orbits}
\label{sec:mwcomp_orbits}

We compare the present-day distances and pericentric distances of the Auriga satellites to those reported for MW satellites in \citet{pace_proper_2022}. The comparison is shown in Figure~\ref{fig:orbit_MW_comp} plotted against stellar mass. Disruption correlates with orbit in the expected sense: the least disrupted satellites ($f_{\rm bound} \gtrsim 0.75$) are found almost exclusively at large present-day distances ($\gtrsim 100$ kpc) and large pericenters ($\gtrsim 60$ kpc), while heavily disrupted satellites extend to much smaller distances and pericenters. The separation is cleaner in pericenter than in present-day distance, consistent with the distance of pericentric passages driving the magnitude of mass loss. The MW satellites occupy the region of this space populated by Auriga satellites spanning a wide range of bound fractions, so orbital properties alone cannot establish that any individual MW satellite is intact.
As described in \citetalias{shipp_auriga_2025}, the distance and pericenter distributions of the Auriga satellites are broadly consistent with those of the MW. However, many simulated satellites on orbits similar to those of MW satellites have experienced high levels of disruption not yet observed among the MW population.

\subsection{Satellite Sizes}
\label{sec:mwcomp_sizes}

\begin{figure*}[t]
    \centering
    \includegraphics[width=0.9\linewidth]{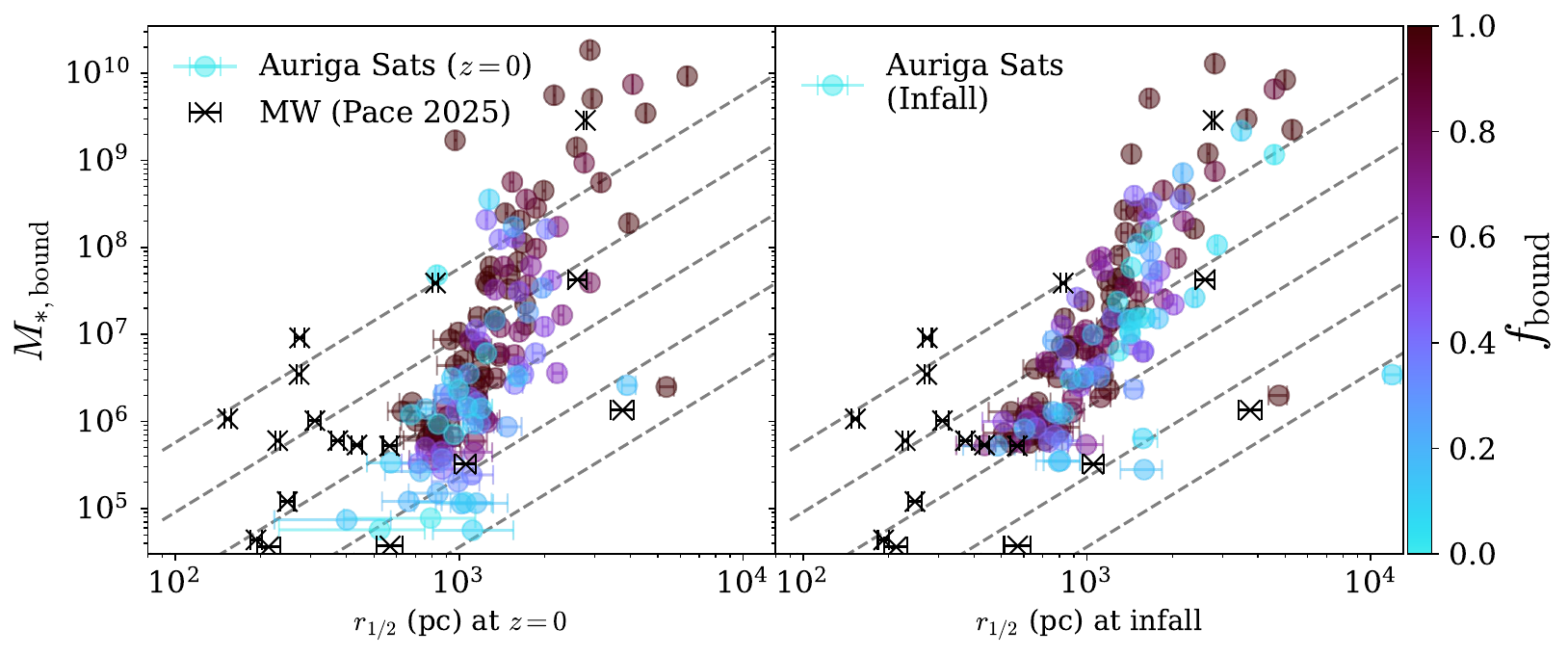}
    \caption{Left: The left panel of Figure~\ref{fig:present_day_m_size_relation} with MW satellites (black crosses) from \citet{pace_local_2025}. Right: The corresponding mass-size relation for Auriga satellites at accretion. At fixed mass, MW satellites are generally smaller than Auriga satellites, and this offset is already present at accretion.}
    \label{fig:mass_size_relation_MW_comp}
\end{figure*}

Although present-day size does not distinguish disrupted from intact satellites in Auriga (Sections~\ref{sec:present_day} and \ref{sec:evo_params}), the mass-size relation remains a direct test of how closely the simulated satellites resemble those of the MW. We compare the 2D projected sizes of Auriga satellites with MW observations from the Local Volume Database \citep{pace_local_2025} in Figure~\ref{fig:mass_size_relation_MW_comp}. The mass-size relation is steeper in Auriga, and at fixed mass, the observed MW satellites are generally smaller than their simulated counterparts. This discrepancy has previously been identified using 3D sizes \citep{bose_cores_2019}, and we find that it persists when comparing analogously measured projected sizes. Importantly, the offset is already present at accretion, before the satellites have experienced significant tidal evolution. Although the mass-size relation at accretion has a slightly shallower slope than at present day, Auriga satellites are systematically larger than MW satellites from the outset, rather than becoming so through tidal evolution after infall.

Several factors could contribute to this discrepancy. Auriga uses the \citet{springel_cosmological_2003} ISM model, which represents unresolved processes in star-forming gas through an effective equation of state \citep{grand_auriga_2017}. The choice of ISM model can strongly affect galaxy sizes \citep{grand_auriga_2017, schaye_colibre_2026}. Numerical resolution may also contribute: gravitational softening and spurious collisional heating can increase simulated stellar sizes \citep{ludlow_numerical_2020, ludlow_spurious_2023}, while the highest-resolution Auriga simulations reproduce the observed sizes of MW and Local Group dwarf galaxies across a wide range of luminosities \citep{grand_determining_2021}. Similarly, extended low-mass satellites are found in other cosmological simulations at comparable resolution \citep{fattahi_tidal_2018}. The present simulations cannot determine the relative contributions of galaxy formation physics and numerical effects to the size offset; we discuss the latter further in Section~\ref{sec:disc_resolution}.

The larger sizes of Auriga satellites could contribute to their greater susceptibility to tidal stripping, since a more extended stellar distribution is generally more vulnerable to tides. However, our results do not suggest that stellar size alone controls the disruption rate. Within Auriga, we find no strong correlation between a satellite's initial size and its subsequent mass loss; orbital history is a much stronger predictor of disruption \citepalias{shipp_auriga_2025}. Moreover, preliminary idealized resimulations with controlled variations in stellar size (V.~Reddy et al., in prep.) find that disruption changes appreciably primarily for extreme size variations. These results suggest that the size discrepancy may contribute to the difference between the simulated and observed populations, but is unlikely to explain it entirely. Nevertheless, the absence of a strong correlation within Auriga does not rule out a systematic effect of the offset between populations, and quantifying its contribution requires controlled comparisons at fixed orbit and dark matter structure.

The inner dark matter distribution introduces a separate, potentially competing effect. Because dark matter dominates the mass of these satellites, their susceptibility to tides depends strongly on their inner dark matter profiles. The Auriga galaxy formation model does not produce cores in its dwarf satellites \citep{bose_cores_2019}, leaving them with cuspy profiles that are relatively resilient to tidal stripping \citep{Read:2006}. In contrast, the inner profiles of classical MW satellites remain debated, with some dynamical analyses favoring cores \citep{walker_slopes_2011} and others finding profiles consistent with cusps \citep{read_heats_2019}. Cored halos lose dark matter more readily to tides than cuspy halos \citep{penarrubia_impact_2010, errani_effect_2017}, and idealized simulations show that this enhanced vulnerability can extend to their stellar components \citep{penarrubia_impact_2010, errani_constraining_2015}. If MW satellites are cored, this difference would act in the opposite direction to the stellar size discrepancy, making them more susceptible to disruption than their cuspy Auriga counterparts. Quantifying these competing effects on cosmological orbits is the subject of ongoing work (V.~Reddy et al., in prep.).

The large stellar sizes and cuspy dark matter profiles of Auriga satellites need not share a common physical origin. In models with bursty stellar feedback, the same processes that expand stellar distributions can also produce dark matter cores, whereas the large sizes in Auriga have been associated with its ISM model \citep{grand_auriga_2017}, which does not produce cores. The combined effects of stellar size and inner dark matter structure therefore cannot be inferred from either discrepancy independently. While both may influence the predicted disruption rate, neither presently provides a complete explanation for the apparent difference between the Auriga and MW satellite populations.

\subsection{Satellite Ellipticities and Orbital Alignments}
\label{sec:mwcomp_ellip}

Another frequently used indicator of tidal disruption is a highly elliptical stellar distribution, which may reflect the elongation of a satellite as it develops tidal tails \citep{oh_tidal_1995, piatek_effect_1995, johnston_interpreting_2002, munoz_turning_2010, kupper_hercules_2017}. \citet{pace_proper_2022} noted that more elliptical MW satellites ($\varepsilon > 0.4$) tend to be elongated along their orbital direction, suggesting that this alignment may reflect tidal tail formation. However, as shown in Section~\ref{sec:evo_params}, Auriga satellites generally become less elliptical after accretion across the full range of bound fractions. Although individual satellites can become elongated during tidal evolution, present-day ellipticity does not reliably trace the amount of mass lost.

\begin{figure*}[t]
    \centering
    \includegraphics[width=1\linewidth]{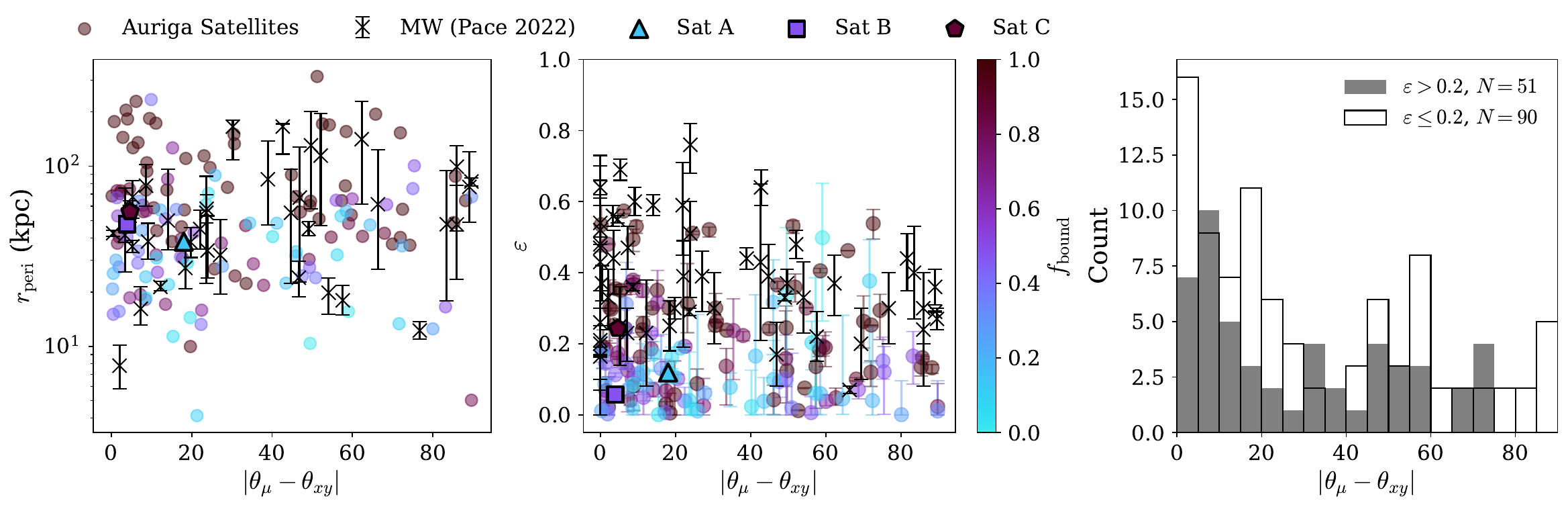}
    \caption{Alignment between proper motion and the direction of elongation, colored by bound fraction. Left: Alignment angle versus most recent pericentric distance from \citetalias{shipp_auriga_2025}. Center: Alignment angle versus ellipticity for Auriga satellites (colored points) and MW satellites (black crosses) from \citet{pace_proper_2022}. Right: Distribution of alignment angles for Auriga satellites divided into high ($\varepsilon > 0.2$) and low ($\varepsilon \leq 0.2$) ellipticity samples. Although \citet{pace_proper_2022} found a preference for alignment among MW satellites with $\varepsilon > 0.4$, Auriga satellites exhibit a preference for alignment across ellipticities and disruption states.}
    \label{fig:ellipticity_peri}
\end{figure*}

Figure~\ref{fig:ellipticity_peri} compares Auriga and MW satellites in the same parameter space as Figure 7 of \citet{pace_proper_2022}. Auriga satellites span a smaller range of ellipticities than MW satellites and are generally more circular; only 12 out of 141 have $\varepsilon > 0.4$. Furthermore, most of the more elliptical Auriga satellites are largely intact: only 8 out of 51 satellites with $\varepsilon > 0.2$ have $f_{\rm bound} \leq 0.5$. Low ellipticity is likewise not a reliable indicator of disruption: 45 out of 90 satellites with $\varepsilon \leq 0.2$ retain $f_{\rm bound} > 0.5$. Thus, satellites spanning a wide range of disruption states overlap substantially in ellipticity.

The difference between the observed and simulated ellipticity distributions may provide clues about differences between the populations. Even largely intact Auriga satellites tend to be more circular than MW satellites, indicating that the discrepancy cannot be attributed to tidal disruption alone. Because Auriga satellites generally become rounder after infall regardless of mass loss (Section~\ref{sec:evo_params}), the more elliptical MW population does not necessarily imply that MW satellites have experienced less tidal processing. Instead, the discrepancy could reflect differences in the shapes of satellites at infall or in their subsequent structural evolution. Spurious collisional heating may contribute to the latter by isotropizing stellar orbits (Section~\ref{sec:disc_resolution}). However, the uncertainties on ellipticities in both the observed and simulated samples are substantial, limiting the significance of this comparison.

The stellar samples used to measure ellipticity also differ. For Auriga, we consider only gravitationally bound stars, whereas observed satellite ellipticities are measured from selected member stars that may include unbound tidal debris. Including such stars could increase the measured ellipticity of observed satellites. However, Appendix~\ref{app:members} shows that the measured ellipticity is relatively insensitive to the stellar sample except for the most disrupted systems, suggesting that this difference is unlikely to fully explain the offset.

\citet{pace_proper_2022} found a preference for alignment between proper motion and the direction of elongation among highly elliptical MW satellites, but not among less elliptical systems, suggesting that the combination of high ellipticity and orbital alignment may indicate tidal disruption. We examine this relationship in \figref{ellipticity_peri}. Although we also find a preference for alignment among highly elliptical Auriga satellites, this preference extends across the simulated population, including satellites with low ellipticity and those that have experienced little mass loss. Alignment therefore does not distinguish strongly disrupted from largely intact systems in Auriga. \citet{yang_shape_2025} similarly found that alignment between elongation and proper motion can arise from projection effects and is a poor indicator of disruption.

Taken together, these results show that neither high ellipticity nor alignment between elongation and proper motion provides a reliable measure of disruption state in the Auriga population. This does not imply that tidal forces cannot produce elongation or alignment in individual satellites, but rather that these observables are not uniquely associated with substantial mass loss. This motivates the alternative diagnostic based on extended stellar distributions developed in Section~\ref{sec:emr}.

\section{Discussion}
\label{sec:discussion}

Our analysis of the Auriga simulations reveals an important limitation in how satellite disruption is commonly inferred from observations: disrupting satellites are structurally self-similar, meaning a satellite that has lost significant mass through tidal stripping is difficult to distinguish from a less massive but intact system. Neither size nor shape provides a reliable disruption signal in a single snapshot. Heavily disrupted satellites follow the same mass-size relation as intact systems, and satellites become more circular, not more elliptical, as they evolve, regardless of whether they lose mass. The apparent alignment between proper motion and elongation is likewise uninformative, and is present across the population regardless of disruption state. Only two properties show a clear correlation with disruption across the population: orbital history, with smaller pericenters and apocenters leading to greater mass loss \citepalias{shipp_auriga_2025}, and the extended stellar distribution, with satellites hosting a larger fraction of stars beyond $3r_{1/2}$ substantially more likely to be disrupted.

Taken together, these results imply that observed satellites thought to be intact are consistent with moderate to significant disruption. This leaves two possibilities, which are not mutually exclusive: either the MW satellites harbor significant hidden disruption, or the simulated satellites are more disrupted than their MW counterparts as a result of physical differences in the galaxy formation model, numerical effects, or both. In the latter case, the simulated and observed populations differ in a way that must be accounted for in any comparison between them, including comparisons restricted to surviving satellites, since many surviving simulated satellites have themselves lost a substantial fraction of their stars (\figref{fbound_hist}). The true answer likely lies between these two limits. We discuss the implications of hidden disruption (Section~\ref{sec:disc_hidden}) and of over-predicted disruption (Section~\ref{sec:disc_nodisruption}), the numerical limitations of the present simulations and the work needed to address them (Section~\ref{sec:disc_resolution}), and the prospects for distinguishing between these scenarios with upcoming observations (Section~\ref{sec:disc_future}).

\subsection{Implications of Hidden Disruption}
\label{sec:disc_hidden}

Satellite galaxies are used as laboratories for galaxy formation and dark matter physics under the implicit assumption that their present-day properties can be mapped to their properties at infall. Tidal evolution alters the observed properties of the population (Sections~\ref{sec:disc_scaling} and \ref{sec:disc_massfunc}), and the presence of energetically unbound stars biases measurements of individual systems (Sections~\ref{sec:disc_dynmass} and \ref{sec:disc_dmdetect}).

\subsubsection{Galaxy Scaling Relations}
\label{sec:disc_scaling}

As satellite galaxies are disrupted, stars in the outskirts are stripped preferentially, altering the observable properties of the bound remnant. This can bias scaling relations built on observed satellite populations in ways that may be difficult to detect without accounting for disruption state.

\citet{riley_auriga_2026} provides an illustration of this mechanism with the mass-metallicity relation: the preferential stripping of older outer stars can increase the mean metallicity of the bound component as disruption proceeds. This effect is visible in systems known to be disrupting: Tucana III sits above the relation \citep{simon_nearest_2017}, and the Sagittarius dwarf spheroidal, with its prominent tidal tails, has a metallicity well above the standard mass-metallicity relation \citep{mucciarelli_abundances_2017}. Similar offsets should be present, at varying amplitudes, in any satellite that has lost outer stars, whether or not its disruption is apparent. If hidden disruption is widespread, it would produce an excess of apparently low-mass, high-metallicity systems, with no accompanying morphological signature to identify the affected satellites.

The deeper concern is that scatter in these relations is itself treated as a physical observable. At low masses, intrinsic scatter in relations such as mass-metallicity and stellar mass-halo mass is traditionally interpreted as evidence for the stochasticity of galaxy formation -- bursty star formation, variation in reionization timing, and halo-to-halo differences in star formation efficiency \citep{kirby_universal_2013, simon_faintest_2019}. The scatter may be driven by more predictable mechanisms such as dark matter accretion and post-reionization quenching \citep{kim_scatter_2026}. However, if a fraction of the observed scatter instead reflects tidal processing at varying amplitude across the population, then inferences about formation physics absorb an environmental systematic.

Hidden disruption similarly complicates comparisons between satellite and field dwarf populations. Differences between the two populations in size, density, or structure are typically attributed to environmental processing or to intrinsic differences in formation. If a substantial fraction of satellites have been tidally stripped, part of any such difference instead reflects tidal evolution, and the satellite population no longer provides a clean sample of low-mass galaxy properties. Because disrupting satellites remain on the same mass-size relation as intact ones (Section~\ref{sec:evo_params}), this contamination cannot be identified or removed system by system; it must instead be forward modeled at the population level.

\subsubsection{Satellite Mass Functions}
\label{sec:disc_massfunc}

A tidally evolved population of satellites has a substantially different stellar mass distribution from its infall distribution, as shown in \figref{mass_pop_func}: the infall population contains many more massive satellites than survive to present day. Because disrupting satellites are structurally indistinguishable from less massive intact systems, the observed mass function cannot be corrected back to the intrinsic one satellite by satellite, and any inference built on it inherits this systematic. This includes the faint end of the galaxy-halo connection, which is constrained almost entirely by the MW satellite system \citep{behroozi_universemachine_2019, nadler_milky_2020}. The models used to fit the MW satellite population account for disruption through the survival probability of each subhalo, but the luminosity of a surviving satellite is set by its peak halo mass and does not evolve, so any stellar mass lost by surviving MW satellites is absorbed into the inferred stellar mass--halo mass relation at low masses. Dark matter constraints are affected in the same way. Warm and self-interacting dark matter models predict fewer low-mass satellites \citep{nadler_effects_2021, tan_mixed_2025}, and a tidally evolved population can mimic this suppression if disruption is not accounted for. Completeness corrections compound the problem, since they assume a size-luminosity distribution for the underlying population to account for the dependence of detectability on surface brightness; if disruption redistributes satellites in that plane, the corrections and every count-based constraint downstream of them inherit the same systematic.

\subsubsection{Dynamical Mass Modeling}
\label{sec:disc_dynmass}

The dynamical masses of dwarf galaxies are inferred from their stellar velocity dispersions under the assumption of dynamical equilibrium, which is violated by tidal disruption. The velocity dispersion of the bound component itself decreases as a satellite loses mass \citep{penarrubia_tidal_2008}. However, the kinematic samples used to measure the velocity dispersions of observed satellites are not restricted to bound stars. Stripped stars remain spatially coincident with the satellite after they become unbound and are difficult to distinguish from bound members, and the inclusion of these stars in the kinematic sample inflates the measured velocity dispersion, leading equilibrium modeling to overestimate the mass of a disrupting system. The magnitude of this bias remains debated: some studies find that equilibrium estimates remain reasonably accurate until the late stages of disruption \citep{tchiorniy_broken_2026}.

Disruption does not bias all mass measurements equally. Because the dark matter halo extends well beyond the stellar component, tidal stripping removes dark matter first: satellites lose the majority of their dark matter before significant stellar mass loss begins \citep{penarrubia_tidal_2008, smith_preferential_2016, errani_constraining_2015}, as seen directly in the dark matter and stellar bound fractions of our example satellites (\figref{stream_properties}). The dark matter that remains bound is concentrated in the region traced by the stars, so dynamical mass estimates within the stellar extent remain approximately valid even for satellites that have lost most of their halo. Extrapolated quantities are far less robust: any satellite showing evidence of stellar stripping has already lost a large majority of its dark matter, so while the measured inner mass remains reliable, extending it to peak or virial quantities can be strongly biased.

These biases feed directly into one of the most active debates in dwarf galaxy dark matter physics: the diversity of central densities among MW satellites and its anti-correlation with pericentric distance \citep{kaplinghat_too_2019, Hayashi:2020}. This trend has been interpreted both as a signature of tidal evolution and as evidence for self-interacting dark matter, in which core formation and gravothermal collapse produce a wide range of central densities \citep{Nishikawa:2020, Correa:2021}. Hidden disruption sits inside this degeneracy: the central densities are dynamical inferences from stellar kinematics, and if satellites at small pericenters carry unrecognized populations of energetically unbound stars, part of the observed density-orbit trend may be measurement bias rather than either tidal or dark matter physics. Attributing the diversity of dwarf densities to dark matter self-interactions requires knowing the tidal processing of each system.

Simulation-based methods offer a path forward. The Graph Neural Posterior Estimator \citep{nguyen_uncovering_2023, nguyen_trial_2025}, trained on simulated satellites from the FIRE-2 suite, reliably recovers peak masses and circular velocities even for systems experiencing tidal effects. However, such methods succeed precisely because their training sets include disrupting systems, so their accuracy is ultimately limited by the fidelity with which the simulations capture tidal evolution, including its numerical convergence.

\subsubsection{Dark Matter Detection}
\label{sec:disc_dmdetect}

Searches for dark matter annihilation target MW satellites because of their high inferred dark matter densities. The expected signal scales with the J-factor, the line-of-sight integral of the squared dark matter density, which is inferred from stellar kinematics through equilibrium modeling \citep{strigari_dark_2018}. The J-factor is dominated by the inner dark matter distribution, which is the last component to be stripped, so the true annihilation signal of a disrupting satellite is relatively robust to the loss of its outer halo. The inferred J-factor is more sensitive, since it is derived from the velocity dispersion and scales steeply with it \citep{pace_scaling_2019}. If the velocity dispersion is inflated by energetically unbound stars, the inferred central density is biased high, overstating both the predicted signal and the constraining power of non-detections \citep{bonnivard_annihilation_2015, bonnivard_contamination_2016}. Disruption can also affect membership selection, since stripped stars have proper motions and line-of-sight velocities similar to those of the bound members and are likely to pass standard membership cuts. The bias introduced by hidden disruption enters through the kinematic measurement rather than through the underlying dark matter distribution, and propagates directly into the exclusion limits on dark matter particle properties derived from satellite observations.

\subsection{What if MW Satellites Are Not Disrupted?}
\label{sec:disc_nodisruption}

The extended stellar distributions of MW satellites are consistent with significant disruption, but this is not the only physically valid interpretation. Extended components could instead reflect accreted stellar halos built from sub-satellites \citep{kadofong_tracing_2020, deason_dwarf_2022, goater_edge_2024, tau_role_2025, boyea_tides_2026} or extended in-situ populations, rather than tidally stripped stars. If improved data show that MW satellites are in fact largely intact, then the simulations systematically over-predict disruption, and the discrepancy identified in \citetalias{riley_auriga_2025} and \citetalias{shipp_auriga_2025} points to the galaxy formation model, the numerical resolution, or both.

If the simulations do over-predict disruption, the consequences extend well beyond satellite counts. Auriga has been used to compare to the MW satellite population \citep{simpson_quenching_2018} as well as to understand the MW's accretion history \citep{vera-casanova_linking_2022, fattahi_origin_2019, deason_eating_2016}. Predictions for stellar halo growth, stream populations, and subhalo survival all inherit the over-prediction, affecting dark matter constraints derived from stream morphology and satellite abundances. 

Several effects could be leading to systematically higher disruption rates. The consequences of the mismatches in stellar size and inner dark matter structure are discussed in Section~\ref{sec:mwcomp_sizes}. Numerical resolution is a third effect (Section~\ref{sec:disc_resolution}), and subhalo finders can additionally lose track of surviving remnants \citep{mansfield_symfind_2024}. Host-to-host variation also matters: halo 6, the Auriga host with the quietest merger history, has a satellite population closely resembling the MW's (Figure~\ref{fig:mass_pop_func}), suggesting that the MW may simply occupy a quiet tail of the host distribution. Relatedly, the MW's specific accretion history may differ from the simulated hosts. A substantial fraction of MW satellites may have been accreted recently or alongside the LMC \citep{kallivayalil_missing_2018, patel_orbital_2020}, significantly altering their tidal evolution.

Distinguishing between hidden disruption in the MW satellites and over-disruption in the simulations is therefore essential for studies of the MW satellite population in the context of \LCDM.

\subsection{Numerical Limitations and Future Simulations}\label{sec:disc_resolution}

The tidal evolution of low-mass satellites is a demanding numerical problem, and fully resolving it remains challenging for cosmological hydrodynamical simulations of MW-mass hosts. This is particularly relevant for the lowest-mass satellites in our sample: a number of satellites, predominantly those with bound dark matter masses below ${\sim}10^{8.8}\,\Msun$, do not satisfy commonly used criteria for resolved tidal evolution based on particle discreteness \citep{van_den_bosch_dark_2018} and force resolution \citep{chiang_universal_2025} (Appendix~\ref{app:resolution}), although both criteria were calibrated on idealized dark-matter-only simulations and their direct application to cosmological hydrodynamical simulations is uncertain. Satellite populations from simulations at this resolution are nonetheless widely used for comparison with the MW, and our results should be read as characterizing these simulated populations.

Numerically enhanced mass loss can allow affected satellites to disrupt on orbits with larger pericenters than they would in a converged simulation, in the same sense as the offset between Auriga and MW streams found in \citetalias{shipp_auriga_2025}. Direct comparisons between resolution levels, however, indicate that this effect is modest. \citetalias{riley_auriga_2025} found that the bound fractions of satellites matched between the level 4, level 3, and level 2 simulations increase only slightly with resolution, and that the excess of streams relative to intact satellites is present at all three resolution levels. Similarly, \citet{pakmor_superstars_2025} found that satellite disruption times in an Auriga simulation with dark matter particles comparable in mass to the star particles are consistent with those in the reference simulation, indicating that the unequal particle masses of the standard runs do not substantially accelerate satellite disruption. Numerical resolution may therefore contribute to the disruption excess, but the available convergence tests suggest that it is unlikely to account for the excess alone; physical differences between the simulated and observed satellite populations are also likely to contribute.

Numerical effects may also shape the stellar structure of the bound component. Galaxy sizes in hydrodynamical simulations are sensitive to the choice of gravitational softening \citep{ludlow_numerical_2020}, and spurious collisional heating by the more massive dark matter particles can inflate and isotropize stellar orbits over time \citep{ludlow_energy_2019, ludlow_spurious_2021}. The strength of this heating depends on both particle number and the ratio of dark matter to stellar particle masses, so it is most significant in lower-resolution, large-volume simulations, but it may still affect the lowest-mass satellites in zoom simulations: when dark matter particles are comparable in mass to the star particles, the lowest-mass Auriga satellites tend toward smaller sizes and are slightly more numerous \citep{pakmor_superstars_2025}. Both effects could contribute to larger simulated sizes at fixed mass and may therefore contribute to the size offset relative to MW satellites (Section~\ref{sec:mwcomp_sizes}), consistent with the agreement between the sizes of Auriga satellites and observed dwarf galaxies at the highest available resolution \citep{grand_determining_2021}. This offset is already present at accretion, before the satellites have experienced significant tidal evolution (Figure~\ref{fig:mass_size_relation_MW_comp}), indicating that it does not arise from numerically enhanced tidal mass loss, although it may still reflect numerical effects in the satellite progenitors. Because spurious collisional heating can continue after infall, it may also contribute to the growth and circularization of satellites after accretion (Section~\ref{sec:evo_params}), and the present simulations cannot fully separate this contribution from that of tidal heating and post-infall relaxation.

These limitations do not affect our results equally. Numerically enhanced mass loss affects the bound fractions of individual satellites, and therefore the relation between bound fraction and pericenter, but direct comparisons between resolution levels find that bound fractions change only modestly with resolution \citepalias{riley_auriga_2025}. We therefore expect the overall degree of mass loss and its dependence on orbit \citepalias{shipp_auriga_2025} to be reasonably robust. Because $\mu_{\rm EMR}$ and $f_{\rm bound}$ are measured from the same star particles, and mass loss of any origin moves stars from the inner to the extended region, we expect the correlation between them to persist in converged simulations, although its calibration may shift; because the extended region is defined in units of $r_{1/2}$, inflated sizes also move the $3$--$10\,r_{1/2}$ annulus to larger physical radii, an effect the present simulations cannot quantify.

The structural trends are more sensitive to resolution, but several features of our results indicate that they are not purely numerical in origin. The part of the size evolution that correlates with bound fraction, the contraction of heavily stripped satellites from their maximum size, is driven by tidal truncation rather than by spurious collisional heating, which tends to inflate stellar systems. This contraction is also the behavior expected from idealized simulations that follow the tidal evolution of satellites at much higher resolution \citep{errani_structure_2022}. Whether the structural self-similarity of disrupting satellites holds for fully converged systems, and how much of the growth and circularization after infall is numerical in origin, remains to be tested.

Importantly, our primary conclusion that present-day size and shape do not reliably distinguish satellites with different disruption states does not depend on interpreting the absolute amount of structural evolution as fully physical. Rather, it follows from the substantial overlap in these observables across satellites spanning a wide range of bound fractions. Establishing the degree to which the detailed evolutionary trajectories are numerically converged is important for understanding their physical origin, but does not by itself imply that size or shape provides a reliable present-day disruption diagnostic.

Taken together, numerical effects could contribute to the tendency for the simulated population to be more disrupted and more extended than a fully converged simulation with the same galaxy formation model would produce, and may explain part of the difference between the Auriga and MW satellite populations. Several complementary efforts will address this. Controlled high-resolution resimulations of satellites on their cosmological orbits, with systematic variations in stellar size, shape, and inner dark matter slope, will separate physical from numerical disruption (V.~Reddy et al., in prep.). Simulations designed specifically to follow the tidal evolution of low-mass satellites, whose requirements differ from those of simulations designed to model the formation of MW-mass galaxies, would benefit from increasing mass and spatial resolution together, ideally with equal-mass dark matter and star particles \citep{ludlow_energy_2019}, since reducing the softening length alone has a limited effect and can itself introduce spurious heating when the mass resolution is too coarse. By suppressing discreteness noise in stripped subhalos \citep{van_den_bosch_dark_2018}, resolving the tidal radius throughout a satellite's orbit \citep{chiang_universal_2025}, and reducing spurious collisional heating, such simulations would follow the tidal evolution of satellites below $M_* = 5\times10^5\,\Msun$ and extend the calibration of $\mu_{\rm EMR}$ into the ultra-faint regime where most observed MW satellites lie. Until such simulations are available, comparisons between simulated and observed satellite populations, including those restricted to surviving satellites, should allow for the possibility that numerical effects contribute to both the disruption and the structure of the simulated satellites.

\subsection{Prospects with Upcoming Surveys}
\label{sec:disc_future}

Distinguishing between a scenario in which simulations over-predict disruption and a scenario in which observations under-detect disruption requires measuring the extended stellar distributions of a large sample of MW satellites. 
To accomplish this, we will need observations that allow us to determine membership for a larger population of stars than is currently possible. 
Current-generation narrow-band photometric surveys such as MAGIC \citep{Chiti_MAGIC_2026} and Pristine \citep{Martin:2024} use photometric metallicities combined with \textit{Gaia} proper motions to efficiently identify candidate extratidal members. 
In the coming years, LSST will provide deep, well-calibrated photometry to improve photometric metallicities \citep{lsst_science_collaboration_lsst_2009}.  
Additionally, proper motions from future data releases of \textit{Gaia} \citep{gaia_collaboration_gaia_2016} and proper motion measurements from both LSST and Roman \citep{McKinnon:2026} will allow us to identify even fainter extratidal members. 
Ultimately, wide-field spectroscopic samples obtained from DESI, 4MOST, PFS, and WEAVE \citep{cooper_overview_2023, de_jong_4most_2019, Chiba:2026, dalton_weave_2012} will provide the best samples with which to search for signatures of disruption. 

Velocity information is the key discriminant. Photometry alone cannot distinguish a tidally stripped envelope from a bound stellar halo, but radial velocities can directly test whether extended stars are energetically unbound. Combining extended stellar distributions with orbital information -- which does correlate with disruption \citepalias{shipp_auriga_2025} -- provides the strongest joint constraint on a satellite's disruption state.

Quantitative predictions will require forward modeling. Mock observations of simulated satellites, in the style of \citet{shipp_streams_2023}, would account for survey-specific completeness and membership selection, enabling measured values of $\mu_{\rm EMR}$ to be translated into posterior constraints on $f_{\rm bound}$, building upon the qualitative consistency arguments presented here. This will be the focus of future work.

\section{Conclusion}
\label{sec:conclusion}

We have used the Auriga suite of cosmological zoom-in simulations to evaluate commonly used indicators of tidal disruption in satellite galaxies, comparing the structural properties of the bound component to the true disruption state across 141 accreted satellites in six MW-mass hosts. Our key findings are:

\begin{enumerate}
    \item \textbf{Disrupting satellites are self-similar, making structural parameters poor indicators of disruption.} Satellites follow a single mass-size relation regardless of disruption state, so a satellite that has lost significant mass through tidal stripping is difficult to distinguish structurally from a less massive but intact system (Figure~\ref{fig:present_day_m_size_relation}). Size and ellipticity do evolve after accretion, but they evolve across the population rather than in proportion to mass loss. The one change that does track disruption, contraction of the most heavily stripped systems, is measured relative to a satellite's accretion time or maximum size and is therefore inaccessible observationally.

    \item \textbf{Satellites become less elliptical, regardless of mass loss.} Satellites circularize after accretion across the full range of bound fractions, so high ellipticity is not a sign of disruption; only 8 out of 51 satellites with $\varepsilon > 0.2$ have $f_{\rm bound} \leq 0.5$. Low ellipticity is not a sign of a disrupted satellite either: 45 of the 90 satellites with $\varepsilon \leq 0.2$ have $f_{\rm bound} > 0.5$ (Figure~\ref{fig:ellipticity_peri}).

    \item \textbf{Proper motion alignment is not a reliable disruption indicator.} Satellites across the full range of bound fractions show a preference for alignment between elongation and proper motion, so this alignment does not reliably distinguish disrupted from intact systems (Figure~\ref{fig:ellipticity_peri}).

    \item \textbf{The extended stellar distribution is correlated with disruption.} Satellites with higher values of $\mu_{\rm EMR}$, meaning more mass between $3r_{1/2}$ and $10r_{1/2}$ than within $r_{1/2}$, are more likely to be disrupted (Figure~\ref{fig:M_ratio_halfrad}, Table~\ref{tab:EMR_fbounds}). The extended stellar distributions measured in a sample of MW satellites occupy the range populated by substantially disrupted Auriga satellites, including among systems traditionally classified as intact, although this comparison does not establish their individual disruption states.
\end{enumerate}

These results suggest that either the MW satellites are more disrupted than has so far been observed, or the simulations are over-disrupting satellites due to the galaxy formation model, numerical resolution, or both. Both effects may contribute, and distinguishing their relative importance will require improved observations and higher-resolution simulations. In either case, the results presented here have important implications for comparisons between simulated and observed satellite populations. The Auriga simulations, like other cosmological simulations at comparable resolution, are widely used to interpret observations of the MW satellite population, and we find that many of their surviving satellites have lost a significant fraction of their stellar mass with no corresponding change in their observable structure. Comparisons between simulated and observed satellite populations therefore need to account for the disruption state of the simulated satellites, regardless of whether the MW satellites are ultimately found to be disrupted. New and upcoming photometric surveys including LSST, Roman, and Euclid, combined with deeper spectroscopy and narrow-band photometry from the next generation of wide-field surveys, as well as higher-resolution simulations that resolve the tidal evolution of satellites in the ultra-faint regime, will dramatically expand our view of the extended stellar distributions of MW satellites and bring us closer to resolving the disruption rate discrepancy between observations and simulations.

\section*{Acknowledgments}
We thank Nondh Panithanpaisal, Tri Nguyen, Phil Mansfield, Ani Chiti, and Guilherme Limberg for enlightening discussions in the preparation of this work. GT thanks Lainey Gammon and Arianna Long for their encouragement and useful comments on an early draft of the manuscript. 

This research used resources of the Argonne Leadership Computing Facility, which is a U.S. Department of Energy Office of Science User Facility operated under contract DE-AC02-06CH11357. 
Financial support for this publication results from grant \#SA-LSST-2024-106b from Research Corporation for Science Advancement. 
This research was supported by a Royalty Research Fund (RRF) grant from the University of Washington. 
A.A. acknowledges support from the Gordon and Betty Moore Foundation. 
The authors acknowledge support from the DiRAC Institute in the Department of Astronomy at the University of Washington. The DiRAC Institute is supported through generous gifts from the Charles and Lisa Simonyi Fund for Arts and Sciences, Janet and Lloyd Frink, and the Washington Research Foundation.
AHR was supported by a fellowship funded by the Wenner Gren Foundation, a Research Fellowship from the Royal Commission for the Exhibition of 1851, and by STFC through grant ST/T000244/1.
RB is supported by the SNSF through the Ambizione Grant PZ00P2\_223532. 
JJ acknowledges support from the National Aeronautics and Space Administration (NASA) under Grant No. 80NSSC25K0365.
RJJG acknowledges support from an STFC Ernest Rutherford Fellowship (ST$/$W003643$/$1). 
FvdV is supported by a Royal Society University Research Fellowship (URF/R/241005).

N.S. used Claude (Anthropic) for assistance with manuscript editing and prose revision. The authors take full responsibility for the scientific content of this paper.

\software{
This work made use of the following software packages: 
\begin{itemize}
\item \texttt{Agama} \citep{vasiliev_agama_2019},
\item \texttt{astropy} \citep{astropy:2013,astropy:2018,astropy:2022,astropy_21262391},
\item \texttt{Cython} \citep{cython:2011},
\item \texttt{Jupyter} \citep{2007CSE.....9c..21P,kluyver2016jupyter},
\item \texttt{h5py} \citep{collette_python_hdf5_2014,h5py_7560547},
\item \texttt{matplotlib} \citep{Hunter:2007},
\item \texttt{nbody\_streams} \citep{nbody_streams}, 
\item \texttt{numpy} \citep{numpy},
\item \texttt{pandas} \citep{mckinney-proc-scipy-2010,pandas_21003741},
\item \texttt{python} \citep{python}, and
\item \texttt{scikit-learn} \citep{scikit-learn,sklearn_api,scikit-learn_20510517},
\item \texttt{scipy} \citep{2020SciPy-NMeth,scipy_20615351}, and 
\item \texttt{ugali} \citep{bechtol_ugali_2015, collaboration_eight_2015}
\end{itemize}
We also thank the maintainers of the arepo-snap-util package.
Parts of the results in this work make use of the colormaps in the CMasher package \citep{2020JOSS....5.2004V,CMasher_14186007}.
Software citation information aggregated using \texttt{\href{https://www.tomwagg.com/software-citation-station/}{The Software Citation Station}} \citep{software-citation-station-paper,software-citation-station-zenodo}.
}
 
\bibliography{main}

\appendix
\restartappendixnumbering

\section{Evolution of Density Profiles: Further Examples}
\label{app:gallery}

\begin{figure}[!htbp]
    \centering
    \includegraphics[width=0.65\textwidth]{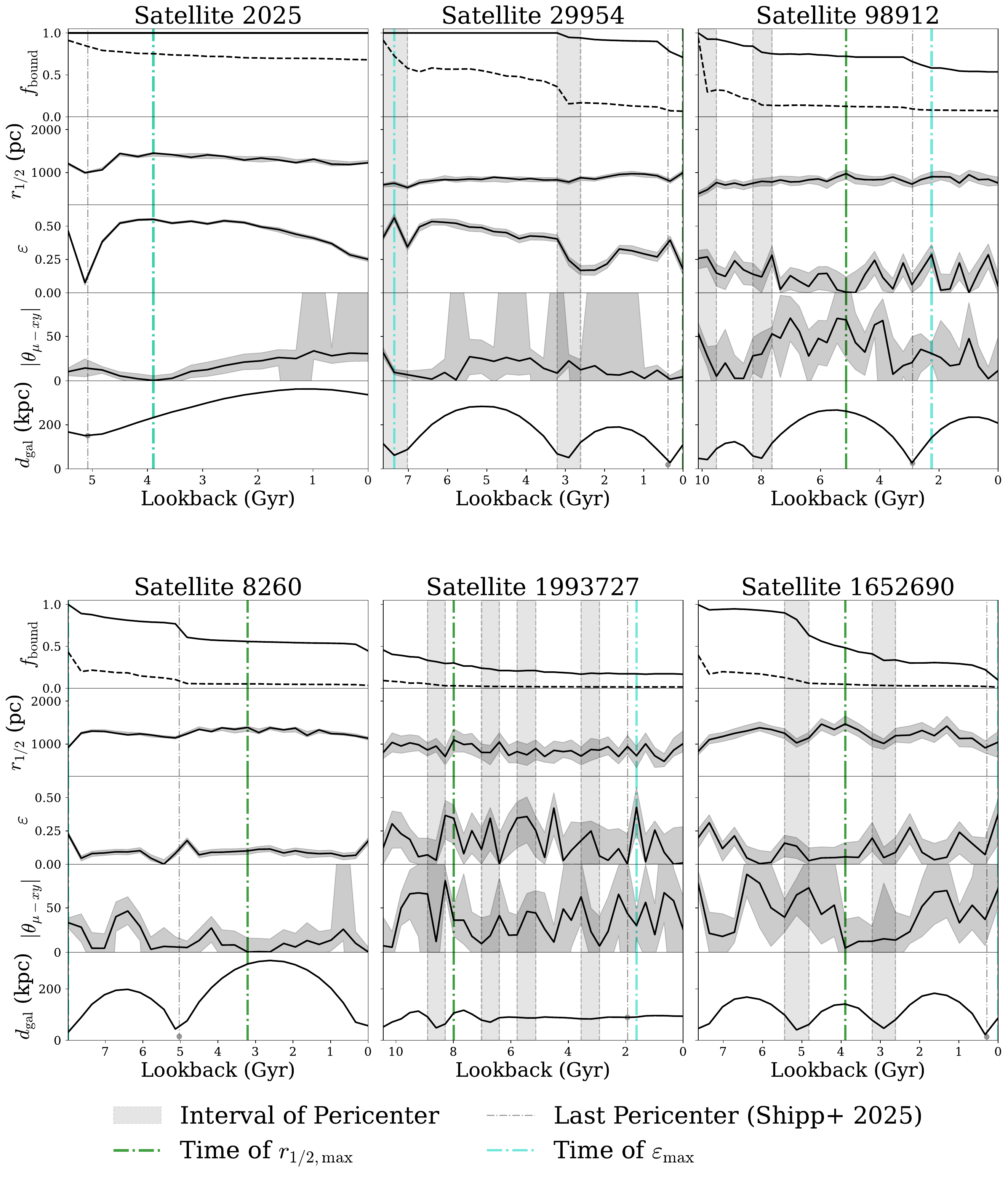}
    \caption{Figure~\ref{fig:stream_properties} for a larger sample of satellite galaxies. Rows are ordered by descending bound fraction at present day. Within each panel, from top to bottom: bound fraction (solid: stellar, dashed: dark matter), half-light radius $r_{1/2}$, ellipticity $\varepsilon$, alignment between proper motion and direction of elongation, and distance from the host center. Markers show simulation output timesteps. Grey shaded vertical regions indicate pericentric passages; the grey dashed line marks the most recent pericenter from \citetalias{shipp_auriga_2025}, with a marker at the pericenter distance. The green dot-dashed line marks the snapshot in which $r_{1/2}$ reaches its maximum; the cyan dot-dashed line marks the snapshot in which the ellipticity reaches its maximum. All of the heavily disrupted satellites have a present-day size below their maximum size, and most also become more circular.}
    \label{fig:example_gallery}
\end{figure}

The 2D elliptical Plummer density profiles used to characterize the satellite galaxies in this work were fit at each snapshot, allowing us to examine the structural evolution of each satellite over time. Figure~\ref{fig:example_gallery} presents a gallery of satellites, with one satellite selected from each of six bins in bound fraction: $f_{\rm bound} > 0.83$, $0.83 > f_{\rm bound} > 0.67$, $0.67 > f_{\rm bound} > 0.5$, $0.5 > f_{\rm bound} > 0.33$, $0.33 > f_{\rm bound} > 0.17$, and $0.17 > f_{\rm bound} > 0.0$. The green dot-dashed line marks the snapshot in which $r_{1/2}$ reaches its maximum, and the cyan dot-dashed line marks the snapshot in which the ellipticity reaches its maximum. These satellites generally follow the same trends as example Satellites A, B, and C. In most cases, the maximum size and ellipticity occur early, followed by a steady decrease to the present day.

\begin{figure}[!htbp]
    \centering
    \includegraphics[width=0.9\textwidth]{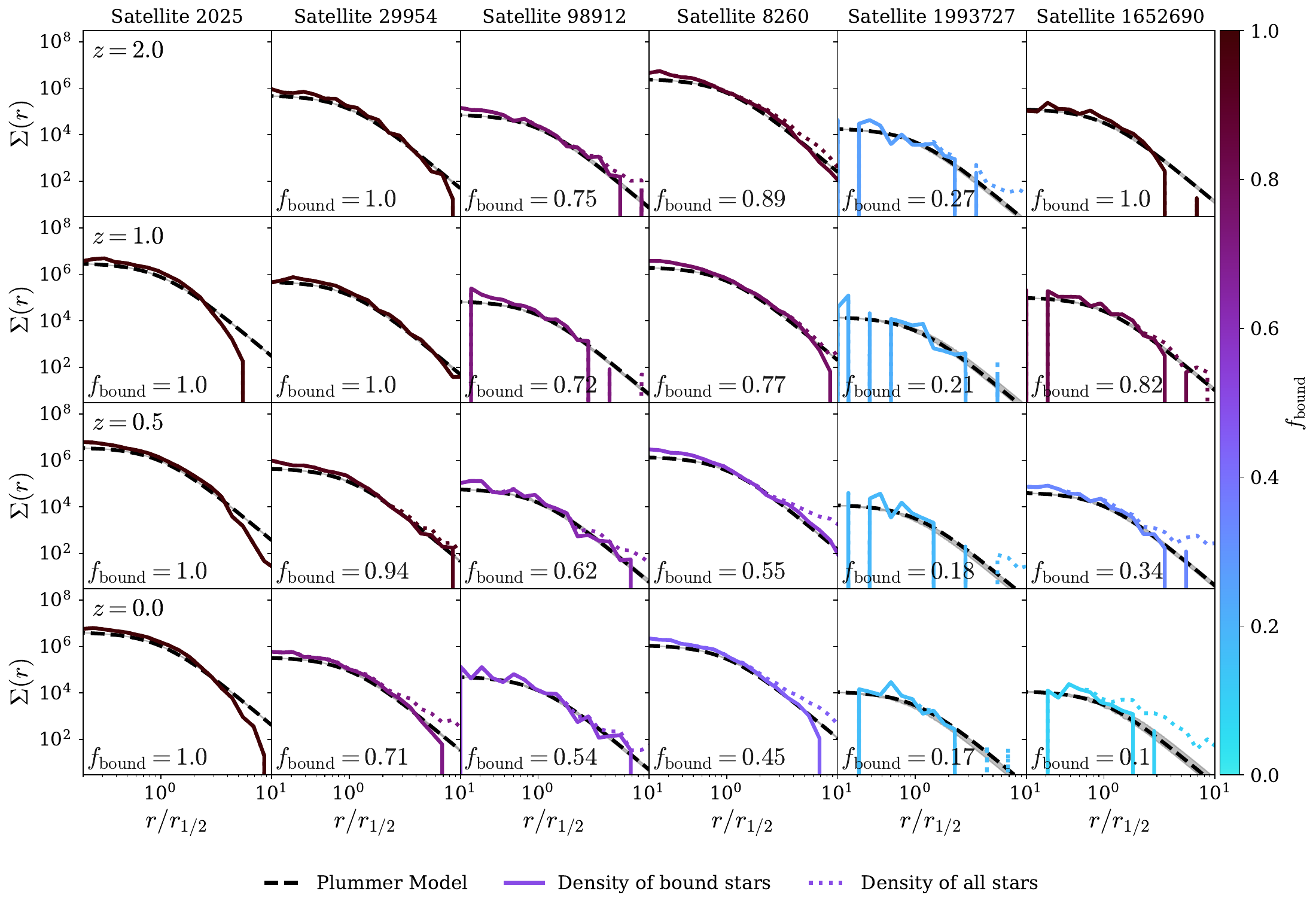}
    \caption{Radially averaged density profiles of the stellar component of the six satellites shown in Figure~\ref{fig:example_gallery}. The solid colored line shows only bound stars, while the dotted colored line shows both bound and unbound stars. The dashed black line is the Plummer model from the fits presented in this work. The model fits the inner regions well, but deviations emerge at $\sim 3r_{1/2}$.}
    \label{fig:density_gallery}
\end{figure}

In Figure~\ref{fig:density_gallery}, we show the evolution of the radially averaged density profiles of the same satellites at $z=2$, $z=1$, $z=0.5$, and $z=0$. The profiles are well fit in the inner regions of each satellite, which contain far more particles and therefore dominate the fit, while the extended regions are poorly fit, especially for the most disrupted satellites. This is expected, as the outer regions are those most affected by tidal disruption and therefore deviate most strongly from a standard density profile.

\begin{figure}[!htbp]
    \centering
    \includegraphics[width=0.75\textwidth]{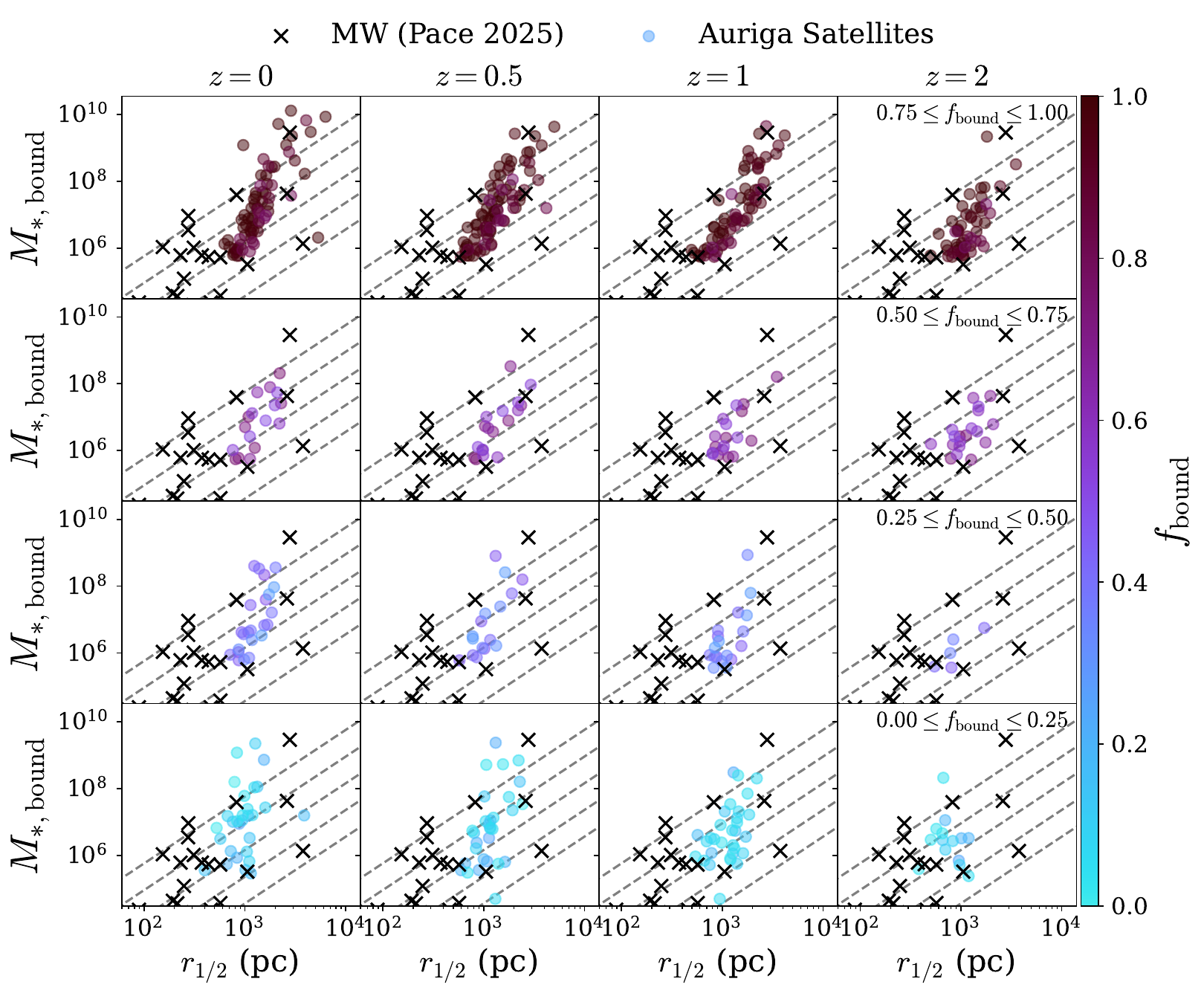}
    \caption{The mass-size relation at (from left to right) $z=0$, $z=0.5$, $z=1$, and $z=2$, divided into four bins in bound fraction (from top to bottom): $1\geq f_{\rm bound}\geq0.75$, $0.75\geq f_{\rm bound}\geq 0.5$, $0.5\geq f_{\rm bound}\geq0.25$, and $0.25\geq f_{\rm bound}\geq0$. Black crosses are MW satellites. The relation remains consistent across time and bound fraction.}
    \label{fig:mass-size_relation_in_time}
\end{figure}

We present the evolution of the mass-size relation at $z=2$, $z=1$, $z=0.5$, and the present day, split into four bins in stellar bound fraction at each redshift, in Figure~\ref{fig:mass-size_relation_in_time}. The relation remains consistent at different times. Comparing the mass and size of satellites across different times, as in Figure~\ref{fig:mass_size_relation_MW_comp}, skews the relation toward the smaller sizes seen at accretion, before the galaxies expand after infall.

\section{Bound Fraction Calculation}
\label{app:fbound}

Throughout this work, we calculate bound fractions using an energy-based criterion implemented in \texttt{nbody\_streams} \citep{nbody_streams}, rather than relying on the \texttt{SUBFIND} catalogs used in previous Auriga Streams papers. We adopt this approach because it can be applied consistently across simulation suites, and because the iterative potential calculation recovers loosely bound material in systems that are actively being stripped, where \texttt{SUBFIND} is most likely to assign these particles to the host \citep{mansfield_symfind_2024}.

\begin{figure}[!htbp]
    \centering
    \includegraphics[width=0.4\textwidth]{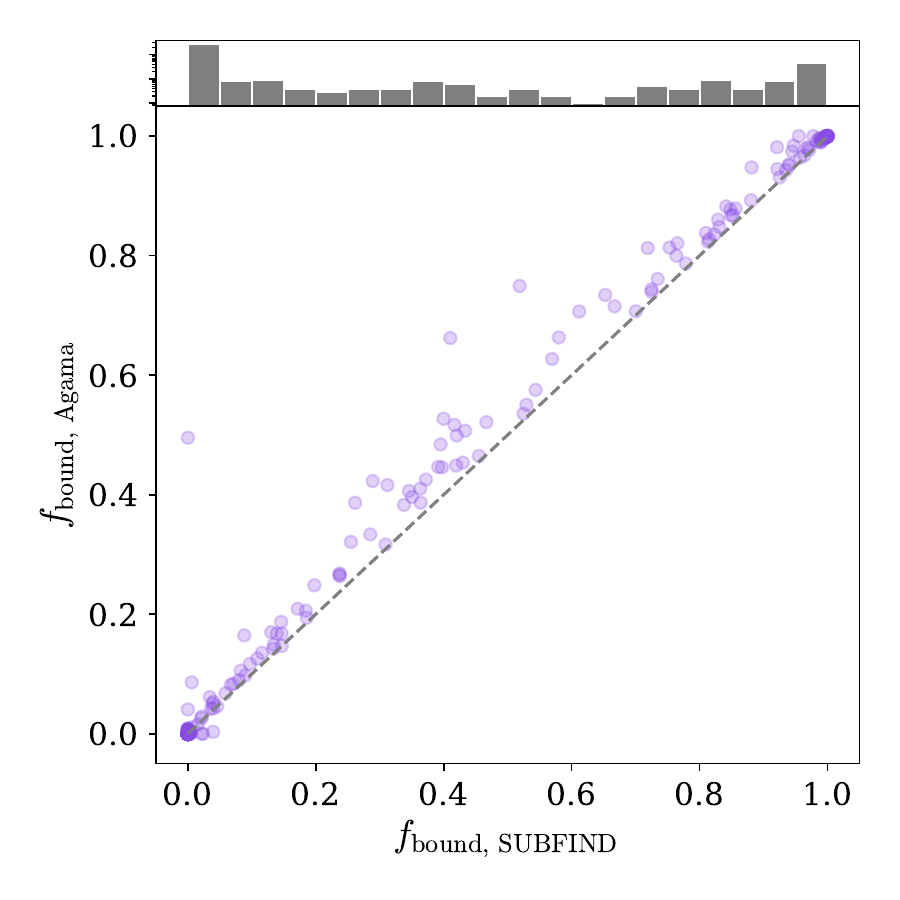}
    \caption{Bound fractions from \texttt{SUBFIND} versus the iterative energy-based calculation with \texttt{nbody\_streams}, for all satellites at all snapshots. The methods agree closely, with the iterative method recovering slightly higher bound fractions for moderately disrupted systems, where \texttt{SUBFIND} is most likely to miss loosely bound material.}
    \label{fig:subfind_agama}
\end{figure}

As shown in Figure~\ref{fig:subfind_agama}, the two methods are in good agreement overall, with the iterative unbinding method recovering slightly higher bound fractions than \texttt{SUBFIND} (median offset $\lesssim 0.002$). The agreement is tightest at the extremes: the most bound and unbound systems ($f_{\rm bound} > 0.8$ and $f_{\rm bound} < 0.2$) show median offsets below $0.001$. Moderately disrupted systems ($0.2 < f_{\rm bound} < 0.8$) show a larger offset, with a median of $\sim 0.007$ and a maximum of $\sim 0.06$, consistent with the iterative unbinding method recovering a more complete bound component in systems undergoing active stripping. The iterative unbinding method also recovers the bound component of one satellite in halo 6 that \texttt{SUBFIND} reported as completely disrupted; this satellite has only 200 star particles, near the lower limit of our mass resolution.

\section{Validation of Energy-Based Membership Criterion}
\label{app:members}

\begin{figure}[!htbp]
    \centering
    \begin{minipage}[b]{0.46\textwidth}
        \centering
        \includegraphics[width=\linewidth]{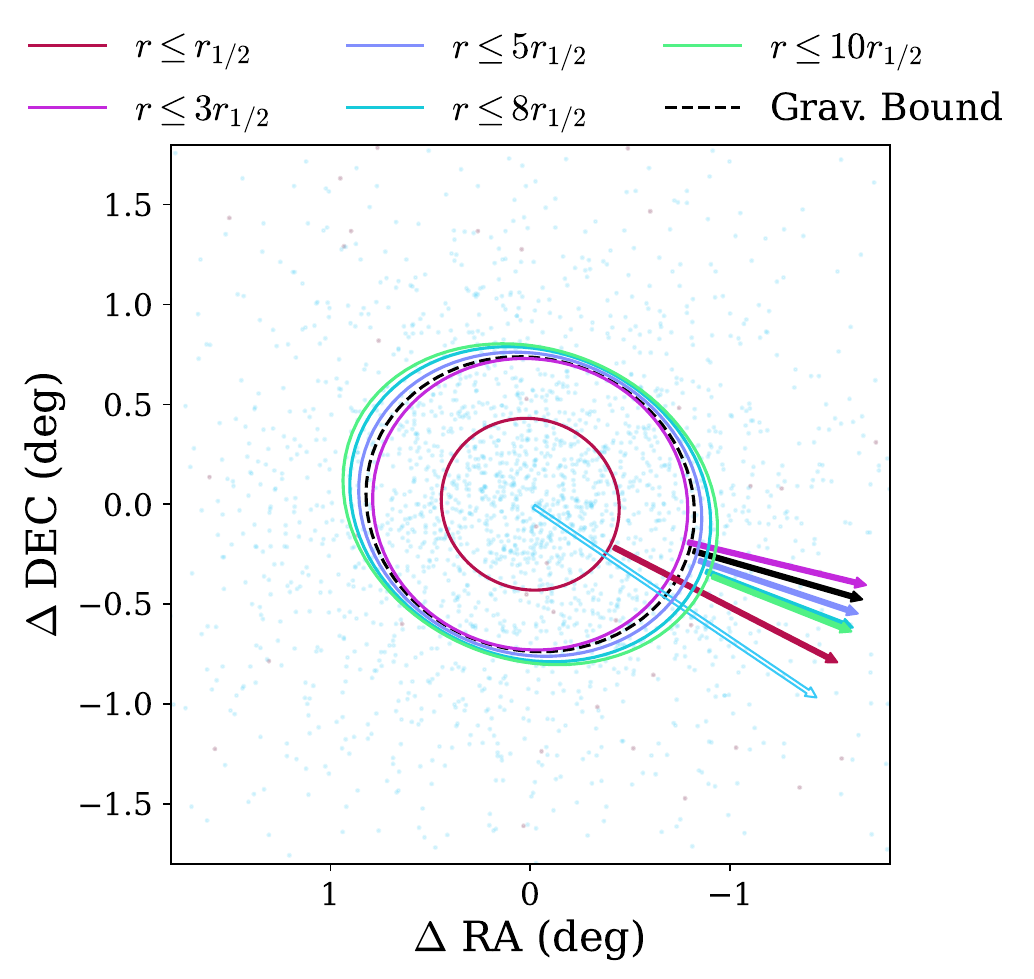}
        \caption{Elliptical Plummer fits to an example satellite using radial membership cuts of 1, 3, 5, 8, and 10\,$r_{1/2}$ (solid colored ellipses), compared to the fit using the energy-based bound membership criterion adopted throughout this work (dashed black ellipse). The blue arrow shows the direction of proper motion; the remaining arrows show the direction of elongation for each fit.}
        \label{fig:example_cuts}
    \end{minipage}\hfill
    \begin{minipage}[b]{0.5\textwidth}
        \centering
        \includegraphics[width=0.9\linewidth]{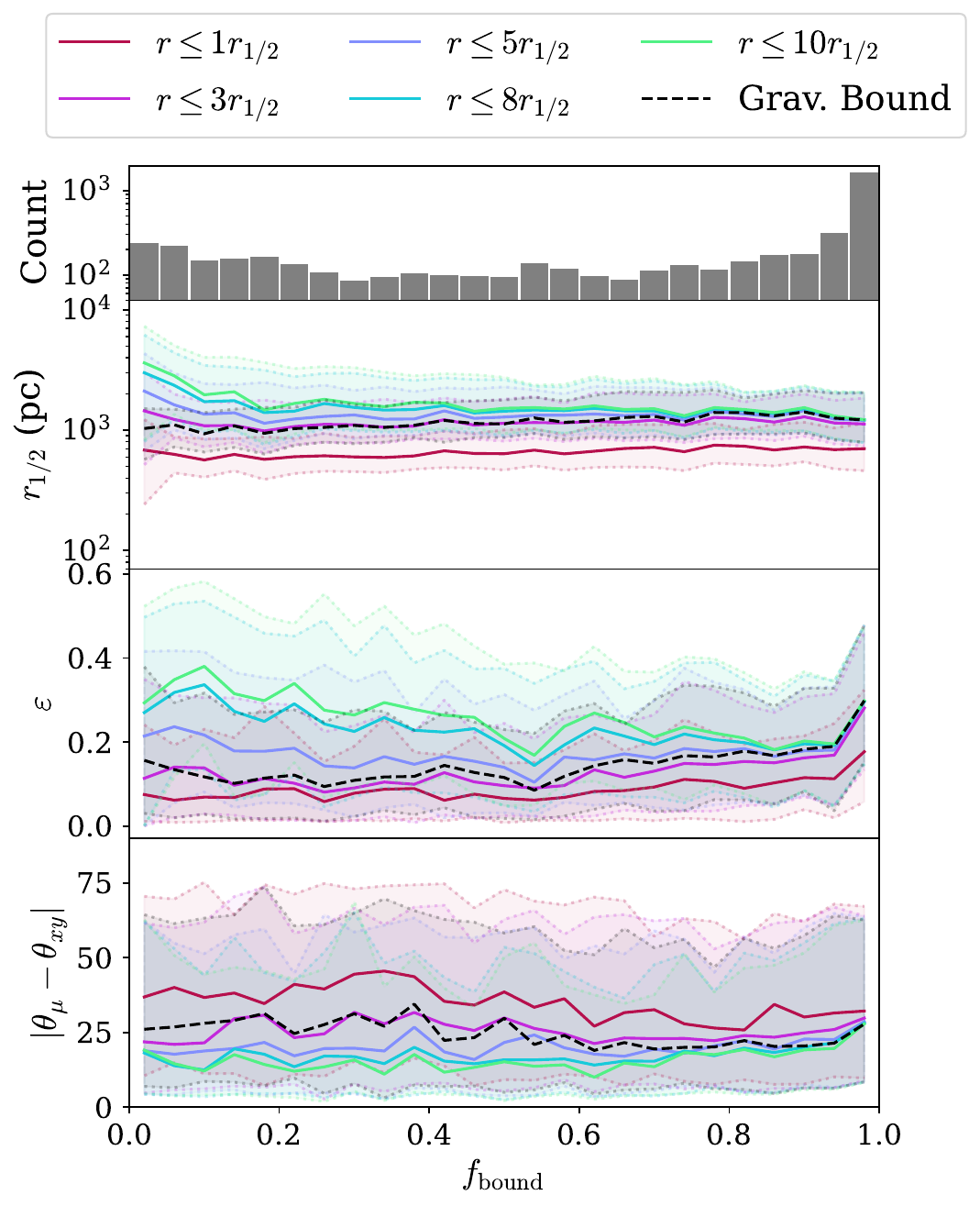}
        \caption{Structural parameters as a function of bound fraction for fits using the energy-based membership criterion and radial cuts of 1, 3, 5, 8, and 10\,$r_{1/2}$. Dashed lines show the median of each parameter across all satellites and snapshots in each bound fraction bin; shaded regions span the 16th to 84th percentiles. The energy-based criterion and the $3r_{1/2}$ cut agree across most of the bound fraction range, diverging only at $f_{\rm bound} \lesssim 0.3$.}
        \label{fig:radialplummers_f_bound}
    \end{minipage}
\end{figure}

To fit the structural parameters of each satellite, we use an energy-based criterion to select only stars that are gravitationally bound to the satellite. This differs from the approach available in observations, where determining whether a star is gravitationally bound is much more difficult than applying a simple radial cut. To ensure that our method produces results consistent with observationally motivated alternatives, we refit all satellites using radial cuts of 1, 3, 5, 8, and 10 times the half-light radius calculated from the gravitationally bound stars. Examples of these fits are shown in Figure~\ref{fig:example_cuts}.

As shown in Figure~\ref{fig:radialplummers_f_bound}, the median structural parameters from the energy criterion are consistent with a radial cut of $R \leq 3r_{1/2}$ across most of the bound fraction range. The two begin to diverge at $f_{\rm bound} \lesssim 0.3$, where the larger radial cuts begin to include stars from the disrupted component. Since our primary focus is on the mostly intact population, where the methods agree well, we adopt the physically motivated energy criterion throughout.

\begin{figure}[!htbp]
    \centering
    \includegraphics[width=0.9\textwidth]{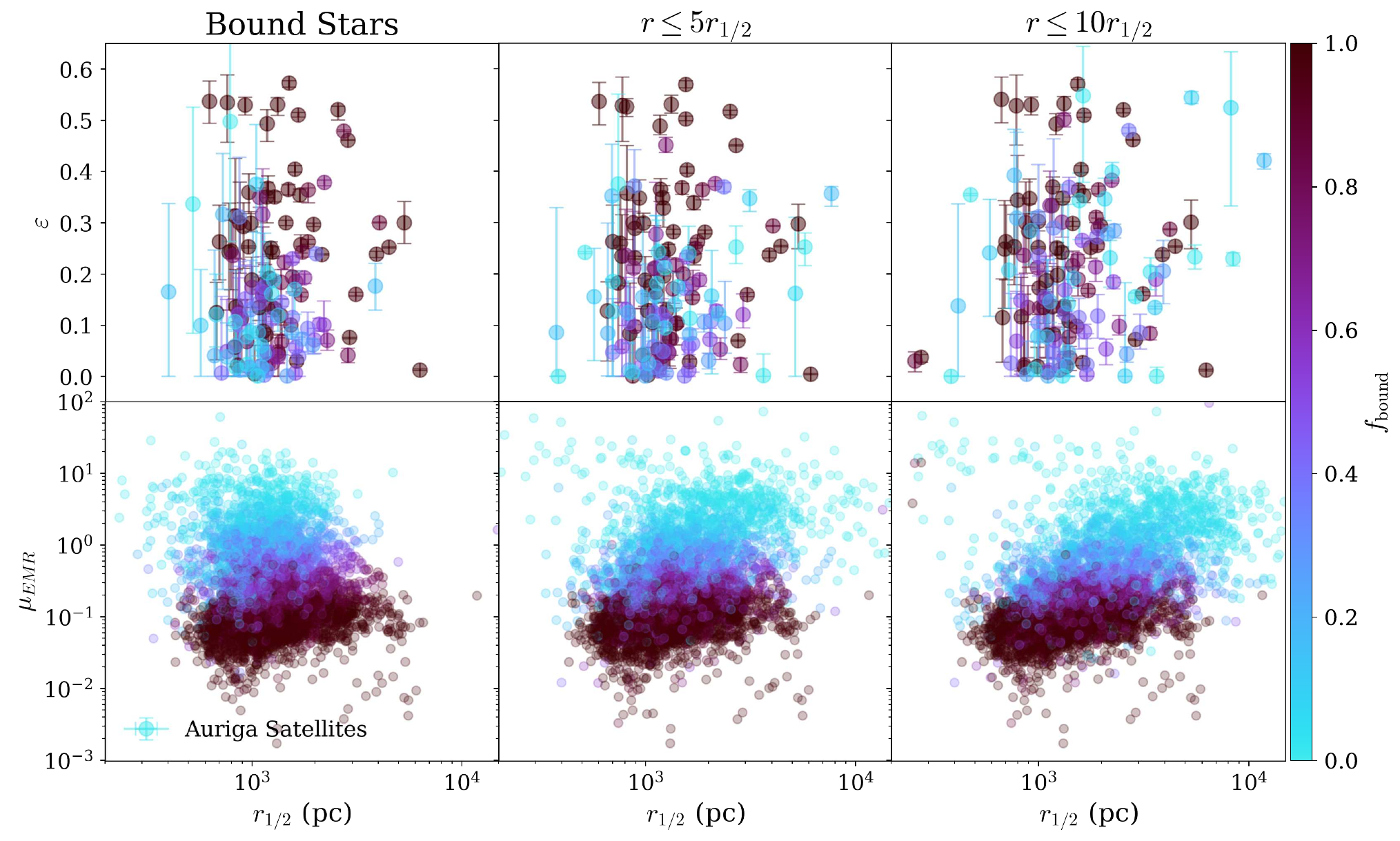}
    \caption{Top: Ellipticity versus half-light radius for fits using the energy-based membership criterion (left), a radial cut of $5r_{1/2}$ (middle), and a radial cut of $10r_{1/2}$ (right); the distribution is insensitive to the choice of membership criterion. Bottom: Extended mass ratio $\mu_{\rm EMR}$ versus half-light radius for the same three membership criteria, colored by bound fraction. Although the larger radial cuts increase the fitted half-light radii of the most disrupted satellites, the correlation between $\mu_{\rm EMR}$ and bound fraction is unchanged.}
    \label{fig:rcut_comp}
\end{figure}

Figure~\ref{fig:rcut_comp} shows that this choice does not affect our conclusions: the distribution of ellipticity versus half-light radius and the correlation between $\mu_{\rm EMR}$ and bound fraction are robust across all membership criteria tested.

\FloatBarrier
\section{Numerical Resolution Criteria}
\label{app:resolution}

To identify satellites whose tidal evolution may not be fully resolved, we apply two criteria that address different numerical effects. The first, from \citet{van_den_bosch_dark_2018}, addresses mass resolution: discreteness noise in subhalos represented by too few particles introduces stochastic scatter in the bound fraction, which can lead either to premature disruption or to spurious survival. They calibrated this criterion using idealized tree-code simulations of isotropic, dark-matter-only NFW subhalos, on circular orbits in a static NFW host potential, and defined it as the bound fraction below which this scatter exceeds 0.1\,dex. The criterion requires the dark matter bound fraction of a subhalo, relative to its mass at accretion, to satisfy
\begin{equation}
    f_{\rm bound,DM} > 0.32\left(\frac{N_{\rm acc}}{1000}\right)^{-0.8},
\end{equation}
where $N_{\rm acc}$ is the number of dark matter particles in the subhalo at accretion. Of the 141 satellites in our present-day sample, 7 (5\%) do not satisfy this criterion.

The second criterion, from \citet{chiang_universal_2025}, addresses force resolution. Using adaptive mesh refinement simulations of dark-matter-only subhalos spanning a wide range of concentrations, velocity anisotropies, and orbital eccentricities in a static NFW host potential, \citet{chiang_universal_2025} find that a subhalo is force-resolved if the minimum tidal radius it has experienced since infall is resolved by at least 20 cells. By comparison with the tree-code criterion of \citet{van_den_bosch_dark_2018}, they find the equivalent condition for tree-based simulations is $r_{\rm t, min} > 20\,\epsilon_{\rm soft}$, where $\epsilon_{\rm soft}$ is the Plummer-equivalent softening length. Unlike the softening criterion of \citet{van_den_bosch_dark_2018}, which they find does not hold for highly eccentric orbits or strongly radially anisotropic subhalos, this criterion is independent of subhalo properties and orbit. For the level 3 Auriga simulations, $\epsilon_{\rm soft} = 188$\,pc at $z<1$ \citep{grand_overview_2024}, corresponding to a threshold of ${\sim}3.8$\,kpc. Since the minimum tidal radius generally occurs at the most recent pericentric passage, we approximate $r_{\rm t, min}$ by the Jacobi radius evaluated at the snapshot closest to each satellite's most recent pericenter,
\begin{equation}
    r_J = \left(\frac{M_{\rm sat}}{3M_{\rm host}(<r_{\rm peri})}\right)^{1/3} r_{\rm peri},
\end{equation}
where $M_{\rm sat}$ is the bound mass of the satellite (dark matter and stars) and $M_{\rm host}(<r_{\rm peri})$ is the host mass enclosed within the pericentric distance $r_{\rm peri}$. This is a simplification of the tidal radius adopted by \citet{chiang_universal_2025}, which uses the satellite mass enclosed within the tidal radius rather than its total bound mass and accounts for the host density profile and the orbital angular velocity of the satellite; the host terms can increase or decrease the tidal radius depending on the orbit, but using the total bound mass systematically overestimates it. Of the 141 satellites in our present-day sample, 75 (53\%) have $r_J < 20\,\epsilon_{\rm soft}$. Because both the use of the total bound mass and the ${\sim}300$\, Myr snapshot cadence, which offsets the evaluated snapshot from the true pericenter, cause $r_J$ to overestimate the minimum tidal radius, this number is likely a lower limit. Nearly all satellites that fail either criterion have bound dark matter masses $M_{\rm DM, bound} \lesssim 10^{8.8}\,\Msun$.

These criteria identify satellites whose tidal evolution may not be fully converged, but they do not measure how strongly any given satellite is affected: failing the force resolution criterion indicates that a satellite's mass loss may be numerically enhanced, and failing the mass resolution criterion indicates that its bound fraction is subject to stochastic scatter in either direction, but neither measures the degree to which its bound fraction or structure differs from a converged result. Direct comparisons of the same halos simulated at different resolutions provide a more direct measure, and find bound fractions that change only modestly between resolution levels \citepalias{riley_auriga_2025}. At the resolution of the level 3 simulations, the satellites that satisfy these criteria are predominantly massive systems on wide orbits that have lost little of their stellar mass, so restricting our analysis to them would not test how the structural trends depend on bound fraction. We use the full sample throughout this work and leave a more detailed characterization of these effects to future work.

\end{document}